\documentclass[12pt]{article}
\usepackage[margin=0.95 in]{geometry}
\usepackage{amsmath} 
\usepackage{slashed}
\usepackage{comment} 
\usepackage{dsfont}
\usepackage{amssymb,amsfonts}
\usepackage[all]{xy}
\usepackage{graphicx}
\usepackage{amsmath}
\usepackage{amssymb}
\usepackage{float}
\usepackage{array}
\usepackage{tikz}
\usepackage{mathtools}
\usepackage{mathrsfs} 
\usepackage[hidelinks]{hyperref}
\usepackage{cite}
\usepackage{tikz}
\usepackage{array, multirow}  
\allowdisplaybreaks
\numberwithin{equation}{section}
\newcommand{\p}{\partial}
\newcommand{\bz}{\bar{z}}
\newcommand{\be}{\begin{equation}}
\newcommand{\ee}{\end{equation}}

 \newcommand{\Tr}{{\text{Tr}}}
\def\bea{\begin{eqnarray}}
\def\eea{\end{eqnarray}}
\numberwithin{equation}{section}
\numberwithin{table}{section}

\begin{document}

\hypersetup{pageanchor=false}
\begin{titlepage}
\vbox{\halign{#\hfil    \cr}}  
\vspace*{15mm}
\begin{center}
{\Large \bf Integrable Structure of Higher Spin CFT \\
and the ODE/IM Correspondence  }

\vspace*{10mm} 

{\large 
Sujay K. Ashok$^{a,b}$, Sanhita Parihar$^{a,b}$, Tanmoy Sengupta$^{a,b}$,\\ 
Adarsh Sudhakar$^{a,b}$, Roberto Tateo$^{c}$ }

\vspace*{8mm}

$^a$The Institute of Mathematical Sciences, \\
		 IV Cross Road, C.I.T. Campus, \\
	 Taramani, Chennai, India 600113

\vspace{.6cm}

$^b$Homi Bhabha National Institute,\\ 
Training School Complex, Anushakti Nagar, \\
Mumbai, India 400094

\vspace{.6cm}

$^c$Dipartimento di Fisica, Università di Torino,\\ 
INFN Sezione di Torino,\\ 
Via P. Giuria 1, 10125, Torino, Italy. 

 \vskip 0.8cm
	{\small
		E-mail:
		\texttt{sashok, sanhitap, tsengupta, adarshsu @imsc.res.in} \\
  \texttt{roberto.tateo@unito.it }
	}
\vspace*{0.8cm}
\end{center}

\begin{abstract} 
We study two dimensional systems with extended conformal symmetry generated by the ${\cal W}_3$ algebra. These are expected to have an infinite number of commuting conserved charges, which we refer to as the quantum Boussinesq charges. We compute the eigenvalues of the quantum Boussinesq charges in both the vacuum and first excited states of the higher spin module through the ODE/IM correspondence. By studying the higher spin conformal field theory on the torus, we also calculate thermal correlators involving the energy-momentum tensor and the spin-3 current by making use of the Zhu recursion relations. By combining these results, we show that it is possible to derive the current densities, whose integrals are the quantum Boussinesq charges. We also evaluate the thermal expectation values of the conserved charges, and show that these are quasi-modular differential operators acting on the character of the higher spin module.  
\end{abstract}

\end{titlepage}

\hypersetup{pageanchor=true}

\setcounter{tocdepth}{2}
\tableofcontents

\section{Introduction and Summary}

Conformal field theory has been a fertile area of research since its inception in the mid-eighties \cite{Belavin:1984vu}, with important connections to critical phenomena, integrable models and string theory \cite{Friedan:1980jm, ZamolodchikovsBook, DiFrancesco:1997nk}. The conformal symmetry in two dimensions is generated by the energy-momentum tensor $T(u)$, whose operator product expansion with itself is equivalent to the Virasoro algebra. In the universal enveloping algebra of the Virasoro algebra, which is generated by $T(u)$ along with the normal ordered composite operators built out of $T(u)$ and its derivatives, it has been shown that there exists an infinite dimensional abelian subalgebra, which is spanned by local integrals of motion \cite{Sasaki:1987mm, Eguchi:1989hs, Kupershmidt:1989bf}.  In a series of papers in the mid-nineties, the problem of diagonalization of these integrals of motion was analyzed \cite{Bazhanov:1994ft, Bazhanov:1996dr, Bazhanov:1996aq, Bazhanov:1998dq}. This problem was referred to as the quantum KdV problem, as quantum integrals of motion reduced to those of the classical KdV problem in the limit of infinite central charge. 

In this work, we study the higher spin version of this problem, which has been referred to in the literature as the quantum Boussinesq problem. We study the higher spin ${\cal W}_3$ algebra, generated by the spin-2 energy-momentum tensor $T(u)$, along with a spin-3 current $W(u)$ \cite{Zamolodchikov:1985wn}. The universal enveloping algebra of the ${\cal W}_3$ algebra is also supposed to contain an infinite dimensional abelian algebra \cite{Kupershmidt:1989bf, Lukyanov:1990tf}. 
Building on the work of \cite{Bazhanov:2001xm} (see also \cite{Fioravanti:1995cq}), we take the next few steps towards finding the infinite-dimensional abelian subalgebra within the universal enveloping algebra of the ${\cal W}_3$ algebra.  
We first provide a brief review of the ${\cal W}_3$-algebra so that the problem we study in this work may be stated more clearly. To begin with, we study the higher spin theory on a cylinder. The spin-2 energy-momentum tensor $T(u)$ and the spin-3 current $W(u)$, have a Fourier expansion of the form
\be
\begin{aligned}
T(u)=-\frac{c}{24}+\sum_{n=-\infty}^{\infty} L_{n}e^{-2\pi i n u},\qquad 
W(u)=\sum_{n=-\infty}^{\infty} W_{n}e^{-2\pi i n u}~.      
\end{aligned}
\ee
Here $u$ is a coordinate on a circle of circumference $1$, and the Fourier expansions imply that we have assumed periodic boundary conditions for the two fields. The modes $L_{n}$ and $W_{n}$ satisfy the $\mathcal{W}_{3}$ algebra \cite{Zamolodchikov:1985wn}: 
\begin{align}
\label{eq:W3 algebra}
[L_n,L_m]  =& (n-m)L_{n+m} + \frac{c}{12}(n^3-n)\delta_{m+n, 0}~, \\
[L_n, W_m] =& (2n-m)W_{n+m} ~, \\
[W_n,W_m] =& \frac{(n-m)}{3}\Lambda_{n+m}+\frac{n-m}{3b^2}\left(\frac{1}{15}(n+m+3)(n+m+2) -\frac16(n+2)(m+2) \right)L_{n+m}\nonumber \\
&+\frac{c}{1080b^2}n(n^2-4)(n^2-1)\delta_{n+m,0}~,  
\end{align}
with $c$ being the central charge and $b^2 = \frac{16}{5c+22}$. The $\Lambda_n$ are the modes of the composite normal ordered operator $(TT)(u)$ built out of the stress tensor: 
\be 
\Lambda_n = \sum_{k=-\infty}^{\infty} :L_kL_{n-k}: + \frac{1}{5}x_n L_n~,
\ee
with $x_{2l} = 1-l^2$ and $x_{2l+1} = 2-l-l^2$. Here, the normal ordering symbol $:\,:$ indicates as usual that we put the operators with larger index $n$ to the right. 
  
In our analysis, we work with just the right-moving or chiral conformal field theory, and the Hilbert space is built out of irreducible highest-weight modules \cite{Bazhanov:2001xm}:
\be 
{\cal H}_{\text{chiral}} = \oplus_a {\cal V}_{\Delta_2^{(a)}, \Delta_3^{(a)}}~.
\ee 
The parameters $\Delta_2$ and $\Delta_3$ are the highest weights. The associated weight-vectors are primaries of the ${\cal W}_3$-algebra and satisfy the following equations: 
\begin{align}
    L_n|\Delta_2, \Delta_3\rangle &= 0 ~, \quad W_n|\Delta_2, \Delta_3\rangle = 0~, \quad \text{for $n>0$}~,\\
    L_0|\Delta_2, \Delta_3\rangle &= \Delta_2\, |\Delta_2, \Delta_3\rangle ~,\quad 
    W_0|\Delta_2, \Delta_3\rangle =\Delta_3\,  |\Delta_2, \Delta_3\rangle ~.
\end{align}
Thus, the pair $(\Delta_2, \Delta_3)$ are the eigenvalues of the operators $(L_0$,$W_0)$  respectively of the highest weight state. 

The universal enveloping algebra of the ${\cal W}_3$ algebra is supposed  to possess an infinite dimensional abelian subalgebra, that is generated by local integrals of motion:
\be 
{\bf I}_k = \int_0^{1}\, du\, J_{k+1}(u)~.
\ee 
The currents $J_{k+1}(u)$ are sums of normal ordered composite operators built out of $T(u)$ and $W(u)$. The first few of these are already written down in \cite{Bazhanov:2001xm} and we list them below:
\be
\begin{aligned}
\label{eq:currents}
    J_{2}(u) &= T(u)~, \\ 
    J_3(u) &= W(u)~,\\
    J_5(u) &= (TW)(u)~,\\
    J_6(5) &= (T(TT))(u) + 9 (WW)(u) + \frac{c-10}{32}(2\pi)^2(T'T')(u)~.
\end{aligned}
\ee
The parentheses indicate conformal normal ordering, and the prime denotes the derivative with respect to the coordinate $u$. The currents are defined up to the addition of total derivatives. 

The expressions for the currents of higher dimensions are not known, and one of our goals is to find a systematic way to construct these currents at higher weights. 
As a first step in that direction, we compute the eigenvalues of the local integrals of motion in the highest weight states:   
\be 
{\bf I_k}|\Delta_2, \Delta_3\rangle = I_k(c,\Delta_2, \Delta_3)|\Delta_2, \Delta_3\rangle~.
\ee 
On the right-hand side, we have explicitly shown the dependence of the eigenvalues not only on the weights but also on the central charge of the higher spin conformal field theory. 
In this work, we compute these eigenvalues for all $k\le 12$, and we do this by exploiting the ODE/IM correspondence for the highest weight state, first observed in \cite{Dorey:1998pt}, and shortly thereafter extended to incorporate general values of the Virasoro vacuum parameter with a proof of the correspondence based on the quantum Wronskian in \cite{Bazhanov:1998wj} (see also \cite{Suzuki:1999rj} for early results on the ODE/IM and  \cite{Dorey:2000ma, Dorey:2006an, Feigin:2007mr} for the higher rank generalization relevant for the present work). We shall also make use of the ODE/IM correspondence to compute excited state eigenvalues, which was first proposed in \cite{Bazhanov:2003ni} for the Virasoro case, with the extension to the higher spin case carried out in \cite{Masoero:2018rel, Masoero:2019wqf}. 

Our analysis in this work rests on the fact that the ODE/IM correspondence\footnote{For reviews of the ODE/IM correspondence, we refer the reader to \cite{Dorey:2007zx, Negro:2017xwc, Dorey:2019ngq}.} provides a link between the spectral theory of ODEs \cite{Sibuya, Voros1, Voros2} and certain integrable quantum field theories, and which relates classical and quantum integrals of motion \cite{Lukyanov:2010rn, Lukyanov:2013wra, Bazhanov:2013cua}. A similar approach to the calculation of the eigenvalues of the integrals of motion in the quantum KdV case has been done in \cite{Bazhanov:1998wj, Dorey:1999uk,  Dymarsky:2022dhi}, and we show how to generalize this to the quantum Boussinesq case. This is one of the main results of this work. 

The vacuum eigenvalues $I_{k}$ turn out to be an important first step in the construction of the currents $J_{k+1}$. In fact, in the quantum KdV case, currents up to weight 12 were derived solely by using the vacuum eigenvalues in combination with an analysis of the conformal field theory on the torus \cite{Maloney:2018hdg}. Following this line of reasoning, we study the higher spin conformal field theory on the torus and compute the one-point function of composite operators ${\cal O}$ built out of the energy-momentum tensor, the spin-3 currents and their derivatives in a higher spin module: 
\be 
\langle {\cal O} \rangle = \text{Tr}_{\text{V}}\, {\cal O}\, q^{L_0-\frac{c}{24}}~,
\ee
with  $q=e^{-\beta}$, and $\beta$ being the inverse temperature, as usual. The states in the module are obtained by acting with the creation operators $L_{-n}$ and $W_{-m}$ on a higher spin primary. 

The thermal one-point functions are computed by first calculating thermal correlation functions involving the energy-momentum tensor and the spin-3 current using the Zhu recursion relations \cite{zhu:1990}, and then by performing conformal normal ordering.  Once these are computed, we propose an ansatz for each of the current densities, as an arbitrary linear combination of composites of $T$, $W$ and their derivatives, with each term having the same conformal dimension. By identifying the low-temperature limit ($q\rightarrow 0$) of the one-point functions of the ansatz with the eigenvalue of the conserved currents in the Virasoro module, we attempt to fix the coefficients appearing in the ansatz.  It turns out that, in the higher spin case, this allows one to fix the currents $J_8$ and $J_9$ up to a single undetermined constant. The current density $J_{10}$ turns out to be a total derivative, while for $J_{11}$, the vacuum eigenvalues fix the current up to four undetermined constants. We check that these results are all completely consistent with the classical limit of large central charge, in which the quantum currents go over to the classical ones. This fixes the large-$c$ behaviour of the undetermined constants. 

Fortuitously, in all these cases, we find that the composite operators multiplying the undetermined constants all have vanishing thermal one-point function in the higher spin module. This allows for an unambiguous calculation of the thermal one-point function of the conserved charges. Moreover we show that these can be written as quasi-modular differential operators acting on the character of the higher spin module. This is the second main result of this work. 

Finally, in the quest to fix the undetermined constants appearing in the current densities, we move on to calculate the eigenvalues of the Boussinesq charges in the first excited level through the ODE/IM correspondence proposed for this case \cite{Masoero:2018rel, Masoero:2019wqf}. Firstly, we find that the sum of the eigenvalues at the first excited level precisely matches the subleading coefficient in the $q$-expansion of the thermal one-point function of the conserved charges, thereby providing a consistency check on our results for the thermal one-point functions. Secondly, we show that by combining the excited state eigenvalues with higher point thermal correlators we are provided with a systematic route to determine the current densities associated with the higher conserved charges. We illustrate this point by a determination of the current density $J_8$.

This paper is organized as follows. In section \ref{classicalBoussinesq}, we review the classical Boussinesq equation, the associated classical conserved charges and identify the classical Lax operator. We also review its relation with the matrix Lax formulation of the modified affine Toda equation and derive the third-order ordinary differential equation that plays a central role in what follows. In section \ref{vacuumODE}, we perform a WKB analysis of the resulting scalar differential operator via the ODE/IM correspondence. We use the asymptotic form of the wave function to read off the eigenvalues of the quantum integrals of motion in the highest-weight modules. In section \ref{thermalcorrelators} we study the higher spin CFT on a torus and compute thermal one-point functions of the conserved charges using the Zhu recursion relations. We show that these agree with the charges of the classical Boussinesq hierarchy in the large central charge limit. As a further check on our results, in Section \ref{excitedODE} we compute the eigenvalues of the conserved charges in the first excited state by working with the appropriate ODE for the excited states and find perfect agreement with the results for the thermal one-point functions. In Section \ref{two_point} we show how the higher point thermal correlators can be used to fix the form of the current density $J_8$. We conclude with a discussion of our results and collect some technical details in the appendices. 

\section{The Boussinesq Equation: Classical Analysis}
\label{classicalBoussinesq}

We review some well-known results related to the classical Boussinesq equation in this section. The Boussinesq equation for $U(x,t)$ is a non-linear partial differential equation given by \cite{boussinesq1872theorie}
\begin{equation}
    \partial_{t}^{2}U=-\frac{1}{3}\partial_{x}^{2}(\partial_{x}^{2}U-2U^{2})~. 
\end{equation}
We begin with the observation that the Boussinesq equation is essentially the consistency condition between the two coupled equations:
\begin{align}
     \p_t U  &= 2 \p_x V \label{eq:Beom1}~,\\
     \p_t V &= -\frac{1}{6}\p_x^3 U + \frac{2}{3} U \p_x U ~. \label{eq:Beom2}
 \end{align}
Furthermore, this problem is integrable, as it can be recast in the form of a Lax equation: 
\be 
\partial_t L = [L,A]~,
\ee 
with the relevant Lax pair given by \cite{Mckean:1978}:
\begin{align}
 L&=\partial_{x}^{3}-U \partial_{x}-\frac{1}{2}\partial_{x}U+V ~,
 \label{eq:SLax}\\
A&= \partial_{x}^{2}-\frac{2}{3}U   ~.
 \end{align}

\subsection{Conserved Charges}
\label{classicalcharges}
The scalar Lax equation is useful to us in the following way: the WKB coefficients from the solution to its eigenvalue problem directly give us the conserved densities. We show this in detail, following the ideas in \cite{zakharov1973stochastization, Mckean:1978}. Consider the scalar Lax equation, and its associated eigenvalue problem:
  \begin{align}
      \p_{t} L &= [A, L]~,  \\
      L \psi &= \lambda \psi~.
  \end{align}
We now differentiate the eigenvalue equation with respect to $t$ and use the Lax equation, keeping in mind its isospectral property. Further, assuming that the span of eigenvectors with eigenvalue $\lambda$ is one dimensional, we arrive at
  \begin{align}
        (L-\lambda) (\p_{t}\psi -A \psi) &= 0~, \nonumber\\ 
        \text{or}\quad (\p_{t}\psi -A \psi) &= g(t) \psi~.
        \label{dtpsieqn}
  \end{align}
Substituting the WKB ansatz $\psi(x, t) = \exp \left(\frac{1}{\epsilon} \int_0^x dx' \ P(x',t,\epsilon)  \right)$ into the $x$-derivative of \eqref{dtpsieqn} gives us:
 \begin{equation}
     \p_{t} P(x,t,\epsilon) - \epsilon~\p_{x}A = 0~.
 \end{equation}
 From this, it follows that the spatial integrals of $P(x,t,\epsilon)$ are conserved quantities. Therefore, we must compute the WKB solutions of the diagonalised Lax operator to find the conserved charges. To this end, we substitute into the eigenvalue problem of \eqref{eq:SLax} the following ansatz:
\begin{equation}
\psi(x,t)=\text{exp}\big(i \frac{x}{\epsilon} +i \int_{0}^{x}\chi(x,t,\epsilon)\big)~.
\end{equation}
For small $\epsilon$ we have the asymptotic expansion:
\begin{equation}
\chi(x,t,\epsilon)=\sum_{n=1}^{\infty}\chi_{n}(x,t)~\epsilon^{n}~.	
\end{equation}
Substituting this into the eigenvalue equation $(L-\lambda)\psi =0$, with $i\lambda = \frac{1}{\epsilon^3}$, we obtain the first two terms in the WKB expansion: 
\begin{align}
\chi_{1}&=-\frac{1}{3}U~,\qquad
\chi_{2}=-\frac{i}{6}U_{x}-\frac{i }{3}V~.
\end{align}
For $n>2$, we obtain a recurrence relation for the $\chi_n$: \begin{align}\label{recrel}
	\chi_{n+1}=i(\chi_{n})_{x}-\sum_{k=1}^{n-1}\chi_{k}\chi_{n-k}+\frac{1}{3}(\chi_{n-1})_{xx}+i\sum_{k=1}^{n-2}\chi_{k}(\chi_{n-k-1})_{x} \nonumber \\ -\frac{1}{3}\sum_{p+m+l=n-1}\chi_{p}\chi_{m}\chi_{l}-\frac{U}{3}\chi_{n-1}~.
\end{align}
By $(\chi_n)_x$ we mean the $x$-derivatives of $\chi_n$. 
The conserved charges are given by 
\begin{equation}
I^{\text{class}}_{n}(x,t)=\int dx~ J_{n+1}(x,t)~,
\end{equation}
where the $J_{n+1}(x,t)$ are proportional to $\chi_n(x,t)$. 
We list the first few non-vanishing charges:
\begin{subequations}
\begin{align}
	I_{1}^{\text{class}} =&\int dx~ U~,\\
	I_{2}^{\text{class}} =&\int dx ~V~,\\
	I_{4}^{\text{class}} =&\int dx ~ U V~,\\
	I_{5}^{\text{class}} =&\int dx~\big( U^3 + 9 V^2 + \frac{3}{4} U_{x}^2\big)~,\\	
 I_{7}^{\text{class}}=&\int dx~\big( U^4+18 V^{2} U+9 V_{x}^2 +\frac{3}{4}U_{xx}^{2}+\frac{9}{2} U U_{x}^{2}\big) \label{I7classical}~,\\
	I_{8}^{\text{class}} =&\int dx~ \big( U^3 V +3 V^3-3 U V U''-\frac{9}{4} V U'^2+\frac{3}{5}U'' V''\big)\label{I8classical}~,\\
 I_{10}^{\text{class}}=&\int dx~\big(U^4 V+6 U V^3+9 V V'^{2}-6 U^{2} V U''-\frac{15}{2}U V U'^{2}+\frac{21}{4}V U''^{2}\notag\\&\hspace{3.8cm}+\frac{15}{2}V U' U^{(3)}+3 U V U^{(4)}+\frac{3}{7}V^{(3)}U^{(3)}\big) ~.
 \end{align}
 \end{subequations}
We observe that all charges $I_{3n}=0$, as the corresponding densities are total derivatives. We have normalized the conserved current densities such that 
\be 
J_{2n} = U^n + \ldots~,\quad\text{and}\quad J_{2n+1} = U^{n-1}\, V + \ldots  
\ee 
One of our aims in this work is to find the conserved currents of the quantum Boussinesq hierarchy. One check on our eventual results will be to compute the classical limit of those currents and match with these classical currents. 

\subsection{The Affine Toda Theory}

In the previous section we showed that the classical conserved charges are encoded in the spatial integrals of the WKB solution of the scalar Lax equation. Following the recent work \cite{Ito:2023zdc}, we shall show how the form of the scalar Lax operator $L$ in \eqref{eq:SLax} arises in the context of the modified affine Toda equation. In that reference, the WKB analysis for the linear problem associated to the modified affine Toda system was used to obtain the classical charges associated to the integrable system. We shall review this and then show that a suitable limit of the Lax operator reduces it to the higher order differential operator that has previously appeared in the context of the ODE/IM correspondence \cite{Dorey:2000ma, Dorey:2006an}. 
Since our goal in the next section will be the WKB analysis of this higher-order ODE in order to derive the quantum conserved charges, we believe this to be a useful detour to establish a path between the classical analysis of this section and the subsequent quantum analysis. 

We begin with the modified affine Toda equation associated to an affine Lie algebra of rank $r$: 
  \begin{equation}
      \p_{\bz}\p_z \phi(z,\bz) - \sum^r_{i>0} \alpha_i \exp\left(\alpha_i \cdot \phi\right) - p(z)\bar{p}(\bz) \alpha_0  \exp\left(\alpha_0 \cdot \phi\right) = 0~.
  \end{equation}
Here, $\phi$ is a Lie algebra valued field, and $\alpha_i$ are the simple roots of the affine Lie algebra $\mathfrak{g}$. This equation  can also be put in the form of a compatibility condition $[\bar{\mathcal{L}},\mathcal{L}]=0$ of two operators $\mathcal{L}$ and $\bar{\mathcal{L}}$:
   \begin{align}
     \mathcal{L} &= \p_z +  \sum_{i=1}^r \p_z \phi_i H_i + \lambda \left( \sum_{i=1}^rE_{\alpha_i}+p(z) E_{\alpha_0} \right) ~,\\ \bar{\mathcal{L}} &= \p_{\bz} + \lambda^{-1} e^{- \sum_{i=1}^r  \phi_i H_i}\left( \sum_{i=1}^rE_{-\alpha_i}+\bar{p}(\bz) E_{\alpha_0} \right) \ e^{ \sum_{i=1}^r  \phi_i H_i} ~,
  \end{align}
where $H_i$, $E_{\alpha_i}$ and  $E_{-\alpha_i}$ are the generators of $\mathfrak{g}$ in the Chevalley basis, and $\phi_i$ are defined through (with $\alpha_i^{\vee}$ being the co-roots):
\begin{equation}
    \phi = \sum_i \alpha_i^{\vee} \ \phi_i(z,\bz)~.
\end{equation}
We now restrict ourselves to the affine algebra $A_2^{(1)}$ and write out the equation $ \mathcal{L} \psi = 0$ explicitly, choosing a particular representation for  the generators (we refer the reader to \cite{Ito:2023zdc} for details):
  \begin{equation}
      \left(\p_z + \begin{pmatrix}
         \p_z \phi_1 & \lambda & 0 \\
         0  & \p_z \phi_2 - \p_z \phi_1 & \lambda \\ \lambda p(z) & 0  & -\p_z \phi_2
      \end{pmatrix} \right) \begin{pmatrix}
          \psi_1 \\ \psi_2  \\ \psi_3 
      \end{pmatrix}= 0 ~.
  \end{equation}
Eliminating $\psi_2$ and  $\psi_3$ using the above equation ($\lambda \neq 0$), we obtain a higher order ordinary differential equation for $\psi_1$:
  \begin{equation} \label{eq:Lax_Linear}
      (-\lambda)^{-3} (\p_z-\p_z \phi_2)(\p_z + \p_z \phi_2-\p_z \phi_1)(\p_z+\p_z \phi_{1}) \psi_1 = p(z) \psi_1 ~.
  \end{equation}
The left hand side of \eqref{eq:Lax_Linear} is in the form of a generalized Miura transform \cite{Drinfeld:1984qv}, where an operator of the form $ \p_z^3 + u_1 \p_z + u_0$ is expressed in factorized form. We can read off the $u_i(z)$ to be:
\begin{align}
    u_1(z) &= \p_z^2 \phi_1 - (\p_z \phi_1)^2 + \frac{1}{2} \p_z \phi_1 \p_z \phi_2 + (\phi_1 \leftrightarrow \phi_2) ~, \\
    u_0 (z) &= \p_z^3\phi_1 + \p_z\phi_1 \left( \p_z\phi_1 \p_z \phi_2 -(\p_z \phi_2)^2- 2\p_z^2 \phi_1 + \p_z^2\phi_2 \right) ~.
\end{align}
Since the scalar Lax operator of the Boussinesq equation is an operator that fits this form, we may equate the coefficient functions with that of \eqref{eq:SLax} to find the map:
\begin{align}
     U(z) &= -\left(\p_z^2 \phi_1 - (\p_z \phi_1)^2 + \frac{1}{2} \p_z \phi_1 \p_z \phi_2 + (\phi_1 \leftrightarrow \phi_2)\right)~, \\ 
     V(z) &=\frac{1}{2}\left( \p_z^3\phi_1 + 2\p_z\phi_1 \left( \p_z\phi_1 \p_z \phi_2 -\p_z^2 \phi_1\right) -\p_z\phi_2 \p_z^2\phi_1  \right) - (\phi_2 \leftrightarrow \phi_1) ~.
\end{align}

\subsubsection{The null conformal limit}

In order to study the integrable structure of the higher spin conformal field theory, we perform a (conformal) double scaling limit, as in \cite{Lukyanov:2010rn}, on the scalar Lax operator associated to the affine Toda system. This essentially corresponds to taking the massless limit of the higher rank integrable system \cite{Ito:2013aea, Adamopoulou:2014fca}. We begin by fixing the asymptotic behaviour of the $\phi_i(z)$ as $z, \bar z \rightarrow 0$. This is obtained by setting 
  \begin{align}
  p(z) &=  s^{3M} -z^{3M}~, \quad\text{and}\quad 
      \phi_i(z,\bz) = \ell_i log \left(z\bz \right) + \mathcal{O}(1)~.
  \end{align}
In addition, we take a conformal double-scaling limit \cite{Lukyanov:2010rn}
 \be 
 \bz \rightarrow 0,\quad  z \sim s \rightarrow 0,\quad \text{and}\quad \lambda \rightarrow \infty, 
 \ee 
 keeping fixed the quantities
  \begin{equation}
      x = \lambda^{\frac{1}{1+M}}z \ , \ E = s^{3M} \lambda^{\frac{3M}{1+M}}~.
  \end{equation}
As a result of taking these limits, the scalar Lax operator reduces to the familiar differential equation associated to the higher spin conformal field theory\footnote{Here we set $\ell_0=\ell_3=0$, but in this way of writing the operator, the generalization to the higher rank affine Lie algebras is evident \cite{Dorey:2000ma, Dorey:2006an}.}  \cite{Dorey:1999pv, Bazhanov:2001xm,  Ito:2023zdc}:
\begin{equation}
       \prod_{j=1}^{3}\left(\frac{d}{dx}+\frac{\ell_j}{x} - \frac{\ell_{j-1}}{x}\right) \psi = (x^{3M} -E)\, \psi ~. 
       \label{scalarLaxthirdorder}
  \end{equation}
Let us summarize what we have reviewed in this section. We started with the linear problem associated to the integrable model that is the modified affine Toda system. What we have shown is that the scalar Lax operator of this integrable system, in the null conformal limit,  reduces to the differential operator familiar from the ODE/IM correspondence for the higher spin conformal field theory. We now turn to the calculation of the spectrum of the quantum conserved charges in a higher spin module of the conformal field theory via the ODE/IM correspondence. 

\section{Eigenvalues from the ODE/IM Correspondence}
\label{vacuumODE}

We have seen in the classical analysis that the WKB solution of the linear problem associated with the Lax operator encodes the classical conserved charges of the Boussinesq equation.
In the quantum case, the action of the charges on a state can be read off from the asymptotic expansion of the logarithm of Baxter's $T$ functions. Recall that in the classical case \cite{Bazhanov:2001xm}:
 \begin{equation}
      \log \{\mathbf{T}(\lambda)\} =  c_0 \ \lambda - \sum_{n>0 } c_n \ I_n^{\text{class}} \ \lambda^{-n}~.
    \end{equation} 
In the context of conformal field theories, these are generalizations of the $T$ functions, where functions appearing in the Lax operator are promoted to fields and products of fields are regularised through the conformal normal ordering \cite{Bazhanov:1994ft}. Via the ODE/IM correspondence, the $T$ functions can be identified with Stokes data, which in turn can be expressed in terms of the Wronskians of subdominant solutions in different Stokes sectors \cite{Dorey:1999uk}. It turns out that the Wronksians of any two solutions to \eqref{eq:Lax_Linear} satisfy the formal adjoint of that equation \cite{Masoero:2019wqf}: 
\begin{equation}
\label{adjointTodaLaxeqn}
     \lambda^{-3}(\p_z+\p_z \phi_{1})(\p_z + \p_z \phi_2-\p_z \phi_1) (\p_z-\p_z \phi_2) \psi_1 = p(z) \psi_1~.
\end{equation}
Therefore, the WKB periods of the adjoint ODE should encode the spectrum of the quantum Boussinesq charges. So we first take the null conformal limit of the adjoint ODE in \eqref{adjointTodaLaxeqn} and perform a WKB analysis of the resulting differential equation. This should encode the eigenvalues of the quantum conserved charges in a highest weight module of the higher spin conformal field theory.  A similar analysis was carried out for the quantum KdV case in \cite{Bazhanov:1998wj, Dorey:1999uk}, in which the differential operator took the form of a Schr\"odinger operator with a $1/x^2$ potential; we now extend the analysis to the third order ODE\footnote{As observed in  \cite{Adamopoulou:2014fca}, the adjoint ODE can also be obtained more directly from the linear system by an alternative scalar reduction.}: 
\be
\left(\frac{d}{dx} - \frac{\ell_1}{x}\right)
\left(\frac{d}{dx} + \frac{\ell_1-\ell_2}{x}\right)
\left(\frac{d}{dx} + \frac{\ell_2}{x}\right)\psi(x) = - ( x^{3M}  - E)\psi(x)~.
\ee

\subsection{The Langer modification and the Modified ODE}

The first task is to transform this equation into a form in which we can use the WKB ansatz, with the role of $\hbar$ being played by $\epsilon = f(E)$, a specific function of the energy $E$. We follow the general logic of \cite{Dorey:2004fk} (see section 6 of that reference, which in turn follows \cite{Langer:1937qr}).
\begin{enumerate}

\item We begin by expanding out the differential equation:
\begin{equation}
\frac{d^3\psi}{dx^3} - \frac{1}{x^2}(\ell_2^2+\ell_1^2+\ell_1+\ell_2 - \ell_1\ell_2) \frac{d\psi}{dx}+\frac{1}{x^3}\ell_2(\ell_1+2)(1+\ell_2-\ell_1)\psi = -(x^{3M}-E)\psi~.
\end{equation}

\item We change coordinates to $x=e^z$, in combination with the redefinition
\be
y = e^{z} \psi ~,
\ee
leading to the equation
\be
\frac{d^3y}{dz^3} -(1+\ell_2^2+\ell_1^2+\ell_1+\ell_2 - \ell_1\ell_2)\frac{dy}{dz} +(1+\ell_1)(1+\ell_2)(\ell_2-\ell_1) y= -e^{3z}(e^{3Mz} -E )y~.
\ee

\item We now scale $z\rightarrow \gamma z$ and see how the differential operator acting on $y$ is mapped  by this scaling. 

\item We then do the inverse of the map in point 1. above, and obtain the equation
\footnotesize
\begin{align}
&\frac{\left(\gamma^3 \left(x^{3 \gamma} \left(x^{3 \gamma M}-E\right)\right)+\left(\gamma +\gamma  \ell_2-1\right) \left(\gamma  \ell_2-\gamma  \ell_1+1\right)
   \left(\gamma +\gamma  \ell_1+1\right)\right)}{x^3}y(x) \nonumber \\
&\hspace{4cm}-\frac{\left(\gamma^2 \left(\ell _1^2-\left(\ell _2-1\right) \ell _1+\ell _2^2+\ell _2+1\right)-1\right)}{x^2}y'(x)+y^{(3)}(x) = 0~.
\end{align}
\normalsize
We choose $\gamma$ such that the coefficient of the $\frac{1}{x^3}$ term is set to zero. This is a cubic equation in $\gamma$, that has three solutions:
\begin{equation} 
\gamma^{(1)}= -\frac{1}{\ell_1+1} ~,\quad\text{or}\quad \gamma^{(2)} = \frac{1}{\ell_2+1}~,\quad\text{or}\quad \gamma^{(3)}= \frac{1}{\ell_1-\ell_2}~.
\end{equation}
We note that there is a complete symmetry between the 3 roots of the su$(3)$ Lie algebra. To make this manifest, we introduce the parameters:
\be
r_1 = \ell_1+1~,\qquad r_2 = -\ell_2-1~, \qquad r_3 = \ell_2-\ell_1~.
\ee
The $r_i$ satisfy the relation $\sum r_i = 0$, and the three solutions for the $\gamma$ that simplify the ODE is given by $\gamma^{(i)} = -\frac{1}{r_i}$. 
\item Without loss of generality, we choose the last of these solutions and obtain 
\be 
y^{(3)}(x)+\frac{r_1r_2}{r_3^2} \frac{ y'(x)}{x^2}   - \frac{1}{r_3^3} ~x^{-\frac{3(r_3+M+1)}{r_3}}~ \left(1-Ex^{\frac{3 M}{r_3}}\right) y(x)= 0~.
\ee
\item Lastly we redefine the coordinate variable and the parameter 
\be 
x= E^{-\frac{r_3}{3M}} t~,\qquad  E= \epsilon^{-\frac{3M}{M+1}}~, 
\ee 
and obtain the final form of the ordinary differential equation whose WKB solution will encode the  eigenvalues of the quantum Boussinesq charges in a highest weight module:
\be
\epsilon ^3\left( y^{(3)}(t) +\frac{r_1r_2}{r_3^2} \frac{ y'(t)}{t^2} \right) -
\frac{1}{r_3^3} t^{-\frac{3}{r_3}(r_3+M+1) } (1-t^{\frac{3M}{r_3}} )\, y(t)= 0~.
\ee

\end{enumerate}
Lastly, we note in passing that we would have obtained the same third order ODE in the $t$-variables if we had begun with the original ODE in \eqref{scalarLaxthirdorder}, but with a different map between the $r_i$ and $\ell_i$ variables. It follows that one can perform the WKB analysis on either the ODE in \eqref{scalarLaxthirdorder} or its adjoint ODE in order to derive the eigenvalues of the local conserved charges in the highest weight state. 

\subsection{WKB Analysis}

We now compute the WKB solutions to the following class of third-order equations (see \cite{Ito:2021boh} for the WKB analysis of a similar third-order ODE but without the single derivative term):
\begin{equation}
     \epsilon^3 \left(\p_t^3 y + \frac{r_1r_2}{r_3^2t^2}\p_t y \right) + p(t) y= 0~.
\end{equation}
For the case at hand, we have 
\be 
\label{pandJ}
p(t) =  -\frac{1}{r_3^3} t^{-\frac{3}{r_3}(r_3+M+1) } (1-t^{\frac{3M}{r_3}} )~.
\ee
To find the WKB solution, we plug in the usual exponential ansatz 
\begin{equation}
   y(t) = \exp \bigg [\int^t dt \ \sum_{i \geq 0}\epsilon^{i-1} a_i(t) \bigg]
\end{equation}
into the differential equation. Collecting the terms at each order in $\epsilon$,  we find, at the first two orders, the equations:
\begin{equation}
\begin{aligned}
     a_0^3 + p(t) &= 0~, \qquad a_1 + \frac{a'_0}{a_0} = 0 ~.
\end{aligned}
\end{equation}
From an analysis of the remaining set of equations we find that the $a_n$, for $n>1$, satisfy the  recursion relation:

\begin{equation}
a_{n-2}'' +     3 \sum_{i=0}^{n-1} a_i  a_{n-i-1}'   + \sum_{i_1+i_2+i_3=n} a_{i_1}a_{i_2}a_{i_3}  + \frac{r_1r_2}{r_3^2t^2}a_{n-2}= 0~,
\end{equation}
with initial conditions 
\begin{align}
    a_0 &= -p(t)^{1/3} ~,\qquad 
    a_1 = - \frac{a'_0}{a_0}~.
\end{align}
%
        
\subsection{Vacuum Eigenvalues from WKB coefficients}

\subsubsection{Period Integrals} 

Once the $a_n$ are computed explicitly in terms of $p(t)$, the conserved charges of the Boussinesq hierarchy can be calculated from the period integrals of the $a_n$. The period integrals in the $t$-plane are equivalent to twice the line integral from $[0,1]$ in the $t$-plane. Much of this analysis is similar to the analogous one carried out for the quantum KdV case \cite{Dorey:2019ngq}. We first define
\be
\label{contourintegral}
\widehat{I}_{n} = \int_0^1\, dt\, a_{n+1}(t) ~.
\ee
We first make a change of variable $t = z^{\frac{r_3}{3M}}~$, and rewrite the integral as 
\be 
\widehat{I}_{n} = \int_0^1\, dz\, S_{n+1}(z) = \frac{r_3}{3M}\int_0^1\, dz\, z^{\frac{r_3}{3M}-1}\, a_{n+1}(z^{\frac{r_3}{3M}}) ~. 
\label{Ihatnaszintegral}
\ee  
This defines the function $S_n(z)$. We can now write a recursion relation for $S_n(z)$, with $n >1$ that follows directly from the recursion satisfied by the $a_n(t)$. The boundary condition for the recursion is given by
\be 
\label{S0expn}
S_0(z) = \frac{1}{3M}~ z^{-1-\frac{M+1}{3M}}~ (1-z)^{\frac13}~,
\ee 
with $S_1(z)$ given by a total derivative\footnote{Note that for $\widehat{I}_{-1}$, we obtain a constant on performing the integral of $S_0(z)$, while we have $\widehat{I}_{0} =0$, since the integrand $S_{-1}(z)$ is a total derivative. So the integrals only lead to non-trivial conserved charges for $n\ge 1$.}. 

We now convert the definite integral in \eqref{Ihatnaszintegral} to an integral over the Pochhammer contour $\Gamma_P$, resulting in 
\be
\widehat{I}_{n} = \frac{1}{(1-m^{(0)}_n)(1-m^{(1)}_n)} \int_{\Gamma_P}~ dz\, S_{n+1}(z)~,
\ee
where $m^{(0)}_n$ and $m^{(1)}_n$ are the monodromy of $S_{n+1}(z)$ around $z=0$ and $z=1$. These monodromies in turn can be calculated directly from the recursion relation satisfied by the $S_n(z)$. A key input for this calculation is the monodromy of $S_0(z)$ about $z=0$ and $z=1$, which can be read off from  \eqref{S0expn}. In addition we also impose trivial monodromy for $S_1(z)$ around $z=0$ and $z=1$, on account of it being proportional to the total derivative of $S_0(z)$. We omit the details of this calculation and present the result (for $n \ge 1$, and for $n\ne 0~\text{mod}~3$):
\be
\label{IhattermsofS}
\widehat{I}_{n} =\frac{1}{(1-e^{-\frac{2\pi i n }{3}})(1-e^{2\pi i \frac{n(1+M)}{3M} } )} \int_{\Gamma_P}~ dz\, S_{n+1}(z)~.
\ee
The change of variables and the recursion relation ensures that the integrand for every $n$ has a linear combination of terms involving only powers of $z$ and $(1-z)$. Thus the integral over the Pochhammer contour turns out to be linear combinations of the Euler beta-function, on account of the integral:
\be 
\int_{\Gamma_P}dz~z^{a-1}(1-z)^{b-1} = (1-e^{2\pi i a})(1-e^{2\pi i b})\, \frac{\Gamma(a)\Gamma(b)}{\Gamma(a+b)}~.
\ee

\subsubsection{Eigenvalues of Conserved Charges}

The main point of this analysis is that the eigenvalues of the conserved charges in the highest weight states, which we denote by $I_n$, are directly proportional to the contour integrals $\widehat{I}_n$ in \eqref{IhattermsofS}. 
Our normalization (see Appendix \ref{ListOfEigenvalues} for more details regarding normalization) is such that the quantum charges have the appropriate classical limit (when $c\rightarrow \infty$). 

From the form of the classical conserved currents, it is not difficult to infer that the first two eigenvalues must correspond respectively to the eigenvalues of the Cartan generators $L_0-\frac{c}{24}$ and $W_0$ in the highest weight modules labeled by $(\Delta_2, \Delta_3)$. These in turn should be proportional to the integrals $\widehat{I}_1$ and $\widehat{I}_2$ respectively. Given that the central charge should be purely $M$-dependent, this motivates the following map between the parameters appearing in the ODE and those in the higher spin conformal field theory\footnote{A comparison with the results in \cite{Bazhanov:2001xm} confirms this proposal. In particular, one can find the map to the parameters appearing in that reference: 
\begin{align}
    g &= \frac{1}{1+M}~,\quad \text{in terms of which}\quad c = 50-24(g+g^{-1})~,\\
    r_1 &= \frac{\sqrt{3}}{g}(\sqrt{3}p_1+p_2)~, \quad r_2 =\frac{\sqrt{3}}{g}(-\sqrt{3}p_1+p_2) ~,
\end{align} 
where $p_1$ and $p_2$ are the momenta associated with the two bosons that correspond to a free field realization of the higher spin conformal field theory in terms of two bosons with background charges.}:
\begin{align}
 \label{cvsM}
         c &= 2- 24 \ \frac{M^2}{M+1}~,\quad 
     \Delta_2 =\frac{\left(r_1^2 + r_1 r_2+  r_2^2\right) }{9(M+1)} -\frac{M^2}{M+1}~,\quad 
      \Delta_3 = -\frac{r_1 r_2 (r_1+r_2)}{27(1+M)^{\frac{3}{2}}}  ~.
\end{align}

Using the inverse map, one can write out the eigenvalues of the higher conserved charges in terms of the conformal field theory data. We find that the charges $I_{3n}=0$, which agrees with the classical analysis. The first few non-vanishing eigenvalues are listed below\footnote{These eigenvalues have also been obtained by computing the higher integrals of motion in the affine Gaudin models \cite{Lacroix:2018fhf, Lacroix:2018itd}. 
}:
\begin{subequations}
   \begin{align}
I_1 & = \Delta_2 - \frac{c}{24} ~,\\
I_2 & = \Delta_3 ~,\\
I_4 &=  \Delta_3 \left(\Delta_2 - \frac{c+6}{24}\right)  ~,\label{I4inlist}\\
I_5 &= \Delta _2^3+9 \Delta _3^2- \frac{1}{8} (c+8) \Delta _2^2+\frac{1}{192} (c+2) (c+15) \Delta _2-\frac{c (c+23) (7 c+30)}{96768} \label{I5inlist}~.
\end{align} 
\end{subequations}
The full list of eigenvalues in the highest weight state of the higher spin module obtained using the ODE/IM correspondence is given in Appendix \ref{ListOfEigenvalues}.

\section{Thermal Correlators}
\label{thermalcorrelators}

In this section, we develop the technology needed to compute the thermal correlators of the conserved charges in the higher spin conformal field theory. This amounts to first computing the thermal correlators of the Boussinesq currents $J_n$. As seen from \eqref{eq:currents}, the currents are linear combinations of normal ordered operators, each of which is made up of the energy-momentum tensor $T(u)$ and the spin $3$ field $W(u)$ and their derivatives. At this point, it is also important to mention that we do not have explicit expressions for the conserved currents $J_n$ for $n>5$. As we shall see, the thermal one-point functions of the composite operators will provide us with a route to derive the conserved current densities. 

The problem of computing the thermal correlators of conserved charges in the quantum KdV case (which only involved the energy-momentum tensor and derivatives) was solved in \cite{Maloney:2018hdg}, and a key element that greatly aided the analysis was that of modular covariance. It was shown that each such correlator could be written as a modular covariant differential operator acting on the Virasoro character. With the higher spin conformal field theory, however, this luxury is not afforded us. Deriving the modular transformation properties of the generalized character $\Tr\Big( q^{L_0-\frac{c}{24}}\, y^{W_0}\Big)$ is still an outstanding problem (see \cite{KanekoZagier, Gaberdiel:2012yb} for some useful remarks in this direction) and we shall instead work with a reduced (higher spin) Virasoro module, setting $y$ to 1. The modular transformation properties of the character,  including insertions of the zero modes of $W_0$ for small values of $n$ have been discussed in detail in \cite{Iles:2013jha, Iles:2014gra} and we shall make extensive use of these results in our analysis. In the higher spin CFT, we shall find that the thermal correlators are quasi-modular differential operators acting on the torus character in the reduced Virasoro module.

Recall that our spatial direction is compact, with period one, $u\sim u+1$. In addition, we compactify the Euclidean time direction as well, with $\tau \sim \tau+\beta$, so that we have a higher spin conformal field theory at a temperature $T=1/\beta$. Thus, we study the higher spin conformal field theory on a Riemann surface with the topology of a torus. As explained in \cite{Dymarsky:2018lhf, Maloney:2018hdg, Maloney:2018yrz}, the correlators of the conserved charges are best understood as the leading terms in a series expansion for the partition function of a generalized Gibbs ensemble, in which chemical potentials are turned on for each of the conserved Boussinseq charges ${\bf I}_k$.  Following \cite{Iles:2013jha, Iles:2014gra}, we shall consider the trace over a restricted module and define the reduced character as 
\be
\chi_{V, \text{red}} = \text{Tr}_V\left( q^{L_0-\frac{c}{24}}\right)~,
\ee
with $q={e^{-\beta}}$, and where $V$ is a highest-weight representation of the higher spin CFT. We shall omit the subscripts in what follows and simply denote the character as $\chi$. The trace is over the restricted Verma module, which is obtained by acting with the creation operators of the $\mathcal{W}_{3}$ algebra, namely $L_{-n}$ and $W_{-n}$ with $n>0$ on a specific primary $|\Delta_2, \Delta_3\rangle $. We write a generic state in the restricted Verma module as
\be
\label{eq:restrictedV module}
\mathcal{P}=\prod_{n>0}(L_{-n}^{a_{n}}W_{-n}^{b_{n}}|\Delta_2, \Delta_3\rangle= \dots (L_{-n}^{a_{n}}W_{-n}^{b_{n}})(L_{-n+1}^{a_{n+1}}W_{-n+1}^{b_{n+1}}) \dots|\Delta_2, \Delta_3\rangle~~\text{where}~a_n,b_n\geq 0~.
\ee
 In this restricted Verma module, for the non-degenerate case, assuming no null vectors for generic values of $c$, $\Delta_2$ and $\Delta_3$, the character $\chi$ is given by:
\be
\begin{aligned}
\label{eq:cha}
  \chi(\tau)= \Tr_V (q^{L_{0}-\frac{c}{24}})
=&q^{\Delta_2-\frac{c}{24}}\prod_{n}\frac{1}{(1-q^{n})^{2}}=\frac{q^{\Delta_2-\frac{c}{24}}}{(\phi(q))^2}~.
\end{aligned}
\ee
In contrast to the Virasoro characters, where the $q$ expansion coefficients give the partition of $n$, in this case, the $q$ expansion coefficients give the square of the partition of $n$ because we have two sets of generators.

The thermal correlator of any composite operator ${\cal O}$ in a highest-weight module $V$ of the higher spin conformal field theory is then defined to be 
\be 
\langle {\cal O} \rangle = \Tr_V \Big
(\mathcal {O} q^{L_0-\frac{c}{24}}\Big)~.
\ee  
We begin with the calculation of the thermal one-point functions of the low-weight currents $J_n(u)$ given in \eqref{eq:currents}. Calculating these traces in the restricted Verma module is challenging due to the non-linear nature of the $\mathcal{W}_3$ algebra. So we use a recursive method due to Zhu to compute the thermal correlators involving the $T(u)$ and $W(u)$ fields. 
We review the basic idea here and refer the reader to Appendix \ref{Apen:Zhu} for a detailed discussion of the Zhu recursion relation. 
We will follow the notations and conventions of \cite{zhu:1990, Gaberdiel:2012yb}, where the $n$-point functions are denoted  
 \begin{equation}
   F((a^{1},z_1),...,(a^{n},z_{n});\tau) := 
z_1^{h_1}\ldots z_n^{h_n}\Tr(V(a^{1},z_1)\ldots V(a^{n},z_n)q^{L_0-\frac{c}{24}})~,
 \end{equation}
where $V(a^i,z_i)$ represents the $i$-th vertex operator, with $a^i$ being the field and $z_i:= e^{2\pi i u_i}$ can be thought of as its location on the plane. For our purposes, the $a^i$ will be either $T$ or $W$. 
 The central idea of the recursion is to expand the first vertex operator $V(a^1, z_1)$ from the $n$ point function in terms of its modes and push the non-zero modes of $V$ around the trace.  As a result, the full $n$ point function can be described as the sum of an $n-1$ point function with one zero mode insertion (coming from the first vertex operator) and terms involving $n-1$-point functions, in which nonzero modes of $V(a^1, z_1)$ act on one of the fields appearing in the $n-1$ point functions.  
 
 Thus, the thermal correlator of the individual terms that appear in the Boussinesq currents can eventually be written as a sum of products of thermal correlators of zero modes of the energy-momentum tensor $\langle (L_0-\frac{c}{24})^n \rangle$ and the spin-3 current $\langle W_0^m\rangle$ on the torus. While the former is easily evaluated, the latter is much more difficult to compute, with explicit results for $m=1,2$ given in \cite{Iles:2013jha, Iles:2014gra}. 

Finally, it is important to note that all the composite operators appearing in the currents are conformally normal ordered. For a composite operator made up of two local fields $A_1(u_1)$ and $A_2(u_2)$, the normal ordered product $(A_1A_2)(u_1)$ is defined using the two-point thermal correlator \cite{Maloney:2018hdg}:
\be
\label{eq:normalorderingcontour}
\langle(A_1A_2)(u_1)\rangle=\frac{1}{2\pi i}\oint_{u_1}\frac{du_2}{u_2-u_1}\langle A_1(u_1)A_2(u_2)\rangle~.
\ee
For a $n$-point correlator involving $n$ such fields, we perform the normal ordering successively from right to left, starting with the normal ordering of $A_n(u_n)$ and $A_{n-1}(u_{n-1})$, and so on, until we end with a composite operator defined at $u_1$. We shall now illustrate these points by explicitly evaluating the thermal one-point functions of the low-weight Boussinesq currents.

\subsection{Thermal One-point Functions}

In this section, we compute the thermal one-point functions of the conserved charges of the quantum Boussinesq hierarchy. One check of these one-point functions is that, in the low-temperature limit, which corresponds to $q= e^{-\beta} \rightarrow 0$, the thermal one-point function of the charges ${\bf I}_k$ should coincide with the eigenvalues $I_k$ computed using the ODE/IM correspondence in the previous section. Of course, except for the first few, the explicit forms of the conserved currents are not known. In our analysis, we show how the evaluation of the thermal one-point functions of composite operators, along with the eigenvalues $I_k$ in the highest-weight states, provide important constraints on the form of the conserved current densities. We shall systematically work our way up the currents in order of increasing weight. 

\paragraph{${\bf I}_1$:} We start with the thermal one-point function  $J_2(u) = T(u)$. The one-point function of a single vertex operator reduces to the evaluation of its zero mode on the torus. We therefore have
\be
\langle T(u)\rangle= F((T)_0;\tau) ~,
\ee
where $(\mathcal{O})_0$ denotes the zero mode of the operator $\mathcal{O}$. 
Since the zero mode of the energy-momentum tensor on the cylinder is $L_0-\frac{c}{24}$, we obtain 
\be
\langle T(u)\rangle=\Tr_{V}\Big(\Big(L_0-\frac{c}{24}\Big)q^{L_{0}-\frac{c}{24}}\Big)=q\frac{\partial}{\partial q}\chi(\tau) ~.
\ee
It is manifestly independent of the $u$ coordinate; thus, doing the $u$-integral is trivial, and we obtain
\begin{align}
\langle {\bf I}_1\rangle = \int_0^1\, du ~\langle T(u)\rangle =q\frac{\partial}{\partial q}\chi(\tau) ~.
\end{align}
In the zero temperature limit, we obtain
\be 
\langle {\bf I}_1\rangle_{q\rightarrow 0} =\Delta_2- \frac{c}{24}~,
\ee
which matches $I_1$, as expected.

\paragraph{${\bf I}_2$:} Next we compute the $\langle J_3(u)\rangle = \langle W(u)\rangle$, for which we get 
\be
\langle J_3(u)\rangle= \langle W_0 \rangle =\Tr_{V}\Big( W_0 q^{L_{0}-\frac{c}{24}}\Big)=\Delta_{3}\chi(\tau)~,
\ee
where we have used the result given in \cite{Iles:2013jha}. The integral over $u$ is trivially done, and we obtain
\be 
\langle {\bf I}_2 \rangle = \Delta_{3}\chi(\tau)~.
\ee 
The low-temperature limit is again trivially taken, and we obtain $I_2=\Delta_3$. 

\paragraph{${\bf I}_4$:}  
For ${\bf I}_4$, we obtain 
\begin{align}
\langle {\bf I}_4\rangle &=\int_0^1\, du~\langle (TW)\rangle(u_1) =\Delta_3\Big( \partial\chi(\tau) -\frac{1}{4}E_2(\tau)\chi(\tau)\Big)~,
\end{align}
where we have defined $\partial = q\frac{\partial}{\partial q}$, and $E_2(\tau)$ is the level two Eisenstein series. We refer the reader to Appendix \ref{app:TW} for a derivation of the one-point function. 
In the low temperature limit, we have $E_2(\tau)\rightarrow 1$ and we therefore obtain 
\be 
\langle {\bf I}_4\rangle_{q\rightarrow 0} = \Delta_3(\Delta_2 -\frac{c}{24}) - \frac{\Delta_3}{4} = \Delta_3\left(\Delta_2-\frac{(c+6)}{24}\right)~, 
\ee
which perfectly matches $I_4$ in \eqref{I4inlist}.
\subsection{Currents and Thermal Expectation Values} \label{thermal_onepoint}
So far we have seen that the thermal one-point functions behave as expected in the low-temperature limit and reproduce the eigenvalues obtained in the previous section. We now illustrate the use of the thermal one-point functions of composite operators in the determination of the conserved currents in the Boussinesq hierarchy. 

\paragraph{${\bf I}_5$:} At weight $6$, the expression for the current is already given in \cite{Bazhanov:2001xm}. We shall rederive this result by using the results for the thermal correlators. We begin with the most general ansatz consistent with the classical current at weight-6:  
\be
\label{J6LC}
\ J_6(u)= \alpha_1\, (T(TT))(u)+\alpha_2\, (T^{\prime}T^{\prime})(u)+\alpha_3\, (WW)(u)~.
\ee
The thermal one-point functions of each of the composite operators have been derived in the appendix. In fact, all we need at this point is the low-temperature limit, and we have
\begin{align}
 \langle(T(TT))\rangle_{q\rightarrow 0} &= 
\Delta _2^3-\frac{1}{8} (c+4) \Delta _2^2+\frac{1}{960} (c+2) (5 c+32) \Delta _2\nonumber \\
&\hspace{5.5cm}-\frac{c \left(35 c^2+462 c+1504\right)}{483840} ~,\\
(2\pi)^2\langle(T^{\prime}T^{\prime})\rangle_{q\rightarrow 0} &=\frac{\Delta _2}{60}-\frac{31 c}{30240} ~, \\
\langle(WW)(z)\rangle_{q\rightarrow 0} &=\Delta_{3}^2-\frac{1}{18}\Delta_{2}^{2}+\Big(\frac{17 c}{3456}+\frac{91}{8640}\Big)\Delta_{2}-\frac{191 c^2}{1741824}-\frac{2101 c}{4354560} ~.
\end{align}
These are independent of $u$, as expected. Demanding that the low-temperature limit of the thermal one-point function of the current in \eqref{J6LC} should reproduce the eigenvalue $I_5$ in \eqref{I5inlist} turns out to uniquely fix the coefficients $\alpha_i$ and we find that 
\be
 J_6(u)=(T(TT))(u)+\frac{c-10}{32}(2\pi)^2(T^{\prime}T^{\prime})(u)+9(WW)(u)~,
\ee
which exactly matches\footnote{\noindent{Note that in our conventions, the circumference of the cylinder is $1$.} } the result for the current in \cite{Bazhanov:2001xm}, and which we quoted in \eqref{eq:currents}. Having completely fixed the coefficients appearing in our ansatz, we present the result for the thermal one-point function of  the conserved charge: 
\begin{multline}
    \langle {\bf I}_5 \rangle = \partial^{3}\chi(\tau)-E_{2}(\tau)\partial^{2}\chi(\tau)+\Big(\frac{c+6}{192}E_{4}(\tau)+\frac{1}{8}E^{2}_{2}(\tau)\Big)\partial\chi(\tau)\\
    -\frac{c(84E_{2}(\tau)E_{4}(\tau)+(66+5c)E_{6}(\tau))}{241920}\chi (\tau )+9\langle W_{0}^{2}\rangle~.
\end{multline}
Here we have chosen to write the thermal one-point function in terms of the expectation values of powers of the zero mode of the spin-3 current, which have been listed in the Appendix (see equation \eqref{vevofW0sq}).

\paragraph{${\bf I}_7$:} The current $J_8$, whose integral gives the conserved charge ${\bf I}_7$, can be written in terms of the following combinations of normal ordered composite operators:
\be 
J_8 = \alpha_1 (T(T(TT)))+\alpha_2(T (T'T')) + \alpha_3 (T''T'')+ \alpha_4 (TWW)+ \alpha_5 (W'W')~.
\ee
Using the low-temperature limit of the one-point functions, and matching to the eigenvalue $I_7$ listed in the Appendix \ref{ListOfEigenvalues}, we find that the current is fixed up to a single constant:
\begin{multline} 
\label{eq:j8unfixed}
J_8 = (T(T(TT))) +18 (T(WW)) +\frac{3}{16}(5c+46)(2\pi)^2  (W'W') \\
-\frac{1}{7680}(5c^2-28c +1124) (2\pi)^4 (T''T'')\\ 
+ (2\pi)^2\, \alpha\, \left( (T(T'T')) - 3 (W'W') + (2\pi)^2 \frac{c-34}{96}(T''T'')\right)~.
\end{multline}
So the low temperature limit alone does not fix the form of the current. Substituting the thermal one-point functions for each of the composites (see Appendix \eqref{app:listofthermalvevsI7}), one can check that a much stronger statement is true:
\be
\left \langle
 (T(T'T')) - 3 (W'W') + (2\pi)^2 \frac{c-34}{96}(T''T'')
 \right\rangle = 0~.
 \ee 
In other words, the thermal one-point function of the operator multiplying the undetermined coefficient $\alpha$ vanishes. Thus, the coefficient $\alpha$ cannot be fixed, even if we compute the trace of the eigenvalues in the higher excited states. However, a happy consequence of this fact is that the thermal one-point function of the charge can be computed without knowing $\alpha$.  Thus, we obtain 
\begin{equation}
    \begin{aligned}
        \langle {\bf I}_7\rangle=&\partial^{4}\chi(\tau)-2E_{2}(\tau)\partial^{3}\chi(\tau)+\Big(\frac{5}{6}E^{2}_{2}(\tau)+\frac{1}{480} (7 c+194)E_{4}(\tau)\Big)\partial^{2}\chi(\tau)\\
        &-\frac{1}{34560}\Big(2400E_{2}^{3}(\tau)+108 (26 + c)E_{2}(\tau)E_{4}(\tau)+(1092 + c (252 + 5 c))E_{6}(\tau)\Big)\partial\chi(\tau)\\
        &+\frac{c}{1105920}\Big(64 E^{2}_{2}(\tau)E_{4}(\tau)+(580 + c (60 + c))E^{2}_{4}(\tau)+256 E_{2}(\tau)E_{6}(\tau)\Big)\chi(\tau)\\
        &+18\Big(q\frac{\partial}{\partial q}-\frac12 E_{2}(\tau)\Big)\langle W_{0}^{2}\rangle~. \nonumber
    \end{aligned}
\end{equation}
In Section \ref{two_point} we shall return to this point of fixing the undetermined parameter $\alpha$. For now, we go on and list the thermal one-point functions of the higher Boussinesq charges.  

\paragraph{${\bf I}_8$:} By similar considerations, we find that the weight nine current $J_9$ can be determined up to a single constant: 
\begin{multline}
    J_9=(T(T(TW)))+3(W(WW))\\
   -\frac{1}{80} (7 c+2)(2\pi)^2 (T(WT'')) -\frac{c (c+59)+390}{1920}(2\pi)^4 (T''W'')\\
    + (2\pi)^2 \, \gamma\,\left((T(WT''))+\frac{5}{2}(W(T'T'))-\frac{1}{24}(c+55)(2\pi)^2 (T''W'') \right)~.
\end{multline}
We refer to Appendix \eqref{app:listofthermalvevsI8} for the results of individual thermal correlators of composite fields present in $J_9$. 
From these results it is easy to check that the operator that multiplies $\gamma$ not only has a vanishing low-temperature limit, but its thermal one-point function also vanishes. Thus, while the current is not determined uniquely by these considerations, it is possible to calculate the thermal one-point function of the conserved charge:
\begin{equation}
    \begin{aligned}
        \langle {\bf I}_8\rangle=&3\langle W_{0}^{3}\rangle+\partial^{3}\langle W_{0}\rangle-\frac{7}{4}E_{2}(\tau)\partial^{2}\langle W_{0}\rangle+\frac{1}{240}(150E^{2}_{2}(\tau)+(75 + 2 c)E_{4}(\tau))\partial\langle W_{0}\rangle\\
        &-\frac{1}{60480}(2730E^{3}_{2}(\tau)+21 (285 + 7 c)E_{2}(\tau)E_{4}(\tau)+(2310 + c (123 + 2 c))E_{6}(\tau))\langle W_{0}\rangle~.  \nonumber
    \end{aligned}
\end{equation}

\paragraph{${\bf I}_{10}$:} The vacuum eigenvalues are much less constraining on the ansatz for ${\bf I}_{10}$ and we find that the current $J_{11}$ is only fixed up to four undetermined constants:
\begin{align}
    &J_{11} =(T(T(T(TW))))+6(W(W(WT)))+(2\pi)^2\frac{3(11c+10)}{16}(W(W^{\prime}W^{\prime}))\cr
    &+ \frac{(2\pi)^4(19 c^2+832 c+1972) }{1536}(T^{\prime\prime\prime}( WT^{\prime}))+\frac{(2\pi)^6(120 c^3+6937 c^2+24719 c-15906)}{483840}(T^{\prime\prime\prime}W^{\prime\prime\prime})\cr
    &+(2\pi)^2\delta_1 \Big( (T(T(WT^{\prime\prime})))+\frac{21}{2}(W(W^{\prime}W^{\prime}))
    +\frac{(2\pi)^2}{192} (19 c+258)(T^{\prime\prime\prime}( WT^{\prime})) \cr
    &\hspace{9cm} +
    (2\pi)^4\frac{(5 c^2+211 c+608)}{2880}(T^{\prime\prime\prime}W^{\prime\prime\prime})\Big) \cr
    &+(2\pi)^4\delta_2 \left((T^{\prime\prime\prime}( WT^{\prime}))+(W(T^{\prime\prime}T^{\prime\prime}))-\frac{5}{6}(2\pi)^2(T^{\prime\prime\prime}W^{\prime\prime\prime})\right)\cr
    &+(2\pi)^4\delta_3 \left((T(WT^{\prime\prime\prime\prime})) +\frac{5}{2}(T^{\prime\prime\prime}( WT^{\prime}))+(2\pi)^2\frac{1}{120} (5 c-21) (T^{\prime\prime\prime}W^{\prime\prime\prime})\right)\cr
    &+(2\pi)^2\delta_4 \Big(T^{\prime}(T^{\prime}(TW)))
    -3(W(W^{\prime}W^{\prime})) -\frac{1}{96}(2\pi)^2(c-150)(T^{\prime\prime\prime}( WT^{\prime})) \cr
    &\hspace{10cm}+ 
    (2\pi)^4\frac{(9 c-185)}{1440}(T^{\prime\prime\prime}W^{\prime\prime\prime})\Big)    
\end{align}
Following the pattern found for the previous two currents, we find that the four combinations multiplying the $\delta_i$ are trace-free. Thus, the thermal one-point function of the conserved charge can be calculated without ambiguity and we find:
\begin{multline}
    \langle {\bf I}_{10}\rangle =\partial^4\langle (W_{0})\rangle-3 E_{2}(\tau)\partial^3\langle (W_{0})\rangle\\
    +6~\partial \langle (W_{0})^3\rangle-\frac{9}{2}E_{2}(\tau)\langle (W_{0})^3\rangle+\Big(\frac{7}{3}E_{2}^{2}(\tau)+\Big(\frac{7}{6}+\frac{c}{48}\Big)E_{4}(\tau)\Big)\partial^2\langle (W_{0})\rangle\\
    -\Big(\frac{13}{24}E_{2}^{3}(\tau)+\Big(\frac{5}{6}+\frac{7c}{480}\Big)E_{2}(\tau)E_{4}(\tau)+\Big(\frac{87}{224}+\frac{193c}{10080}+\frac{5 c^2}{24192}\Big)E_{6}(\tau)\Big)\partial\langle (W_{0})\rangle\\
    +\frac{1}{3870720} \Big(8 \left(25 c^2+2428 c+53700\right) E_6(\tau ) E_2(\tau )
    +5 \left(c^3+126 c^2+4972 c+72504\right) E^2_4(\tau )\\
    +448 (17 c+945) E_4(\tau ) E^2_2(\tau )+107520 E^4_2(\tau )\Big) \langle W_{0}\rangle~.
\end{multline}

\subsection{The Classical Limit} \label{classical_limit}

In the large central charge limit, the quantum Boussinesq charges should go over to the classical conserved charges derived in Section \ref{classicalcharges}. As we shall see, this imposes some constraints on the form of the undetermined coefficients $\alpha$, $\gamma$ and $\delta_i$. The way to take the classical limit has already been outlined in \cite{Bazhanov:2001xm}, and we review this now. We first rescale and redefine
\be 
T(u)\longrightarrow \left(\frac{c}{24}\right) U(u)~,\qquad W(u) \longrightarrow i \left(-\frac{c}{24}\right)^{\frac32} V(u) ~.
\ee
Then, the quantum conserved charges go over to the classical ones in the following manner: 
\be 
{\bf I}_k \longrightarrow i^{-k-1}\left(-\frac{c}{24}\right)^{\frac{k+1}{2}}\, I_{k}^{\text{class}}~, 
\ee
in the limit that $c\rightarrow -\infty$. Doing this for the first three charges is a trivial exercise, as there is only a single term in both the classical and quantum currents. For ${\bf I}_5$, it has been checked already in \cite{Bazhanov:2001xm} that the classical limit gives the expected classical current. 

We now turn to the constraints that follow from taking the classical limit for the higher Boussinesq currents. We find a perfect agreement with the classical charges, provided we set 
\begin{equation}
    \begin{aligned} 
\alpha &=\phantom{-} \frac{3c}{16} + \alpha^{(0)} ~,\hspace{2.3cm}
\gamma = -\frac{3c}{32} + \gamma^{(0)} ~,\\
\delta_1 &=-\frac{c}{4} + \delta_1^{(0)} ~,\hspace{2.5cm}
\delta_2 =\phantom{-}\frac{7c^2}{768} + \delta_2^{(1)}c + \delta_2^{(0)}~,\\
\delta_3 &=\phantom{-}\frac{c^2}{192} + \delta_3^{(1)}c + \delta_3^{(0)} ~,\qquad
\delta_4 = -\frac{5c}{16} + \delta_1^{(0)}~.
    \end{aligned}
    \label{largeclimitofcoefficients}
\end{equation}    
Thus the leading-in-$c$ behaviour of the various constants is uniquely fixed by the classical limit\footnote{The fact that $\delta_2$ and $\delta_3$ are both quadratic polynomials of the central charge follows by requiring that the operators multiplying these contribute in the classical limit. A linear polynomial in $c$ would lead to a vanishing contribution in each case, and there would be a mismatch with the classical limit.}. We emphasize here that $\alpha^{(0)}$, $\gamma^{(0)}$ and $\delta_i^{(j)}$ are $c$-independent  numbers. Requiring a consistent classical limit of the proposed conserved quantum currents is a non-trivial constraint since the constants $\alpha$, $\gamma$ and, $\delta_i$ feed into multiple terms that appear in the classical conserved charges.  Thus, it is also a useful consistency check, both of the expressions for the currents and of our results for the thermal correlators.

\subsection{Discussion}

We conclude this section with a few remarks about the nature of our results. We have calculated the thermal one-point functions of all Boussinesq charges up to weight $10$. We have expressed these as differential operators acting on the character (with insertions of the zero mode $W_0$), and with coefficients that are quasi-modular forms. As shown in \cite{Iles:2013jha, Iles:2014gra} for $n=1, 2$, the thermal expectation values $\langle W_0^n \rangle$ can also be computed as quasi-modular differential operators acting on the character. Thus, the thermal one-point functions can be expressed as quasi-modular differential operators acting on the character of the higher spin module. This is expected from a study of the generalized Gibbs ensemble, in which one defines a partition function that includes chemical potentials for all the higher spin conserved currents \cite{Dijkgraaf:1996iy}. The thermal correlators we have defined should be thought of as the lowest-order terms in a power series expansion (in the fugacities) of the generalized partition function. 

Secondly, we emphasize the fact that we could calculate the thermal one-point functions even without fixing the form of the current densities exactly. The currents $J_8$, $J_9$ and $J_{11}$ have undetermined parameters, but these did not appear in the thermal one-point functions. This is on account of certain trace-free combinations appearing in the current densities. 

As a first step towards the determination of the undetermined constants, we took the classical limit of large central charge and fixed the leading coefficients in the large-$c$ limit in equation \eqref{largeclimitofcoefficients}. To completely fix the remaining parameters, we need more data from the conformal field theory side. For this purpose, we now calculate the eigenvalues of the conserved currents in the first level of excited states in the higher spin module. This will prove to be useful in two ways: first of all, we should match the sum of the eigenvalues (of the conserved currents) in the first excited level with the terms that appear at sub-leading order in the $q$-expansion of the thermal one-point functions. This is indeed what we shall find in our analysis in Section \ref{excitedODE}. Secondly, as we shall show in Section \ref{two_point}, the excited state eigenvalues open up the possibility of fixing the undetermined parameters of the current densities via two-point functions involving the conserved currents. 

\section{Excited State Eigenvalues from ODE/IM}
\label{excitedODE}

In this section we first review the results of \cite{Masoero:2019wqf} in which they have proposed the ODE corresponding to higher excited states of the quantum Boussinesq model. The third order ODE that is considered in \cite{Masoero:2019wqf}  is given by
\begin{equation}\label{beq:ode1}
	\phi ^{(3)}(z)-W_1(z) \phi '(z)+W_2(z) \phi (z)=0~,
\end{equation}
where the coefficient functions are given by
\begin{eqnarray}
W_{1}&=&\frac{\bar{r}_1}{z^2}+\sum _{j=1}^N \left(\frac{a^{(j)}_{11}}{z \left(z-w_j\right)}+\frac{3}{\left(z-w_j\right){}^2}\right) ~,\\
W_{2}&=&\frac{\bar{r}_2}{z^3}+\frac{1}{z^2}+\lambda 
z^k+\sum _{j=1}^N \left(\frac{a^{(j)}_{22}}{z^2 \left(z-w_j\right)}+\frac{a^{(j)}_{21}}{z \left(z-w_j\right){}^2}+\frac{3}{\left(z-w_j\right){}^3}\right)~.
\end{eqnarray} 
Here $N$ is the level of excited state; for $N=0$, the ODE should therefore be mapped to the one we studied in Section \ref{vacuumODE}. We will show this explicitly in the following subsection. For now, we simply state that the parameters $(\bar r_1, \bar r_2)$ will be mapped to $(\ell_1, \ell_2)$ variables, while $k$ and $\lambda$ will be expressed as functions of $M$ and $E$ (see equation \eqref{mapofparameters} for the map of parameters).   

We shall set $N=1$ in the rest of this section. For this case there are four undetermined coefficients $(a_{11}, a_{21}, a_{22}, w_1)$ in the excited state ODE, and these are fixed by imposing the constraint that the monodromy about $w_1$ is trivial\footnote{From the Bethe ansatz and $TQ$ approach to the integrable structure of the conformal field theory, this condition guarantees that the $Q$ function corresponding to the excited state (whose eigenvalue we compute using the ODE) has the same asymptotics as the vacuum solution \cite{Bazhanov:2003ni}. See \cite{Conti:2021xzr} for a proof of this claim for the quantum KdV models associated with simply laced Lie algebras.}. These will lead to exactly four algebraic constraints that completely fix these parameters. 
The imposition of the flatness conditions is carried out in detail in \cite{Masoero:2019wqf,Masoero:2018rel}, and we only present the results\footnote{We have corrected some minor typographical errors in the results of \cite{Masoero:2019wqf}.}. The coefficients $a_{11}, a_{22}$ and the location of the pole $w_1$ are completely determined in terms of $a_{21}$, as follows:
\begin{align}
a_{11} &= k ~,\\
a_{22}&=\frac{2k+3}{3}a_{21}-\frac{k^{2}}{3} ~,\\
w_1 &= \frac{1}{9(k+2)}
\Big(2 a_{21} (3 + k) ((k+1)(k+3)- \bar{r}_{1})\nonumber \\ 
&\hspace{5cm}-(3 + k) (k (k+1) (k+3)-(k+9) r_1+9 r_2)\Big)~.
\end{align}
The coefficient $a_{21}$ is in turn determined by the quadratic equation: 
\begin{align}\label{quad_levelone}
(a_{21})^2 - k a_{21}  + k (3 + k) - 3 \bar{r}_{1}&=0~.	
\end{align}
Putting all this together and relabelling $a_{21} = a_1$, we have the ODE corresponding to the first excited states of the higher spin module:
 \begin{align}
 \label{excODEv1}
	\left(\frac{\bar{r}_2}{z^3}+\frac{1}{z^2}+\lambda 
z^k+ \left(\frac{(2k+3)a_1-k^2}{3z^2 \left(z-w_1\right)}+\frac{a_1}{z \left(z-w_1\right){}^2}+\frac{3}{\left(z-w_1\right){}^3}\right)\right) \phi (z)\nonumber\\-\left(\frac{\bar{r}_1}{z^2}+ \left(\frac{k}{z \left(z-w_1\right)}+\frac{3}{\left(z-w_1\right){}^2}\right)\right) \phi '(z)+\phi ^{(3)}(z)=0~.
\end{align} 

\subsection{WKB Analysis}

As a first step towards performing the WKB analysis for the excited state ODE, we make a variable change to map the differential equation in \eqref{excODEv1} as a perturbation of the vacuum ODE in \eqref{scalarLaxthirdorder}. We make the following change of parameters and variables:
\begin{equation}
\begin{aligned}
        k &= -\frac{3M+2}{M+1}~,\quad \lambda= (-1)^{\frac{-1}{(M+1)}}(3(M+1))^{-\frac{3 M}{M+1}}E,\\
    \bar r_1 &= -1 + \frac{\ell_1(\ell_1+1)+\ell_2(\ell_2+1)+1-\ell_1\ell_2}{9(M+1)^2}~,\\
    \bar r_2 &= ~\frac{(\ell_1-3M-2)(\ell_2+3M+4)(\ell_1-l_2+3+3M)}{27(1+M)^3}~,\\
    z &= -\frac{x^{3(M+1)}}{27(1+M)^3}~,\qquad \phi(z) = x^{3M+2}\, \psi(x) ~.    
\end{aligned}
\label{mapofparameters}
\end{equation}
It is then straightforward to check that the excited state ODE takes the form
\begin{multline}
    \frac{d^3\psi}{dx^3} - \frac{1}{x^2}(\ell_2^2+\ell_1^2+\ell_1+\ell_2 - \ell_1\ell_2) \frac{d\psi}{dx}+\frac{1}{x^3}\ell_1(\ell_2+2)(1+\ell_1-\ell_2)\psi +(E-x^{3M})\psi\\
    -\frac{9 (M+1) x^{3 M+1} \left(x^{3 M+3}-27 (M+1)^3 (3 M+2) w_1\right) }{\left(x^{3 M+3}+27 (M+1)^3 w_1\right){}^2}\psi '(x) \\
    +\frac{9 a_1 (M+1)^2 x^{3 M}\left(2 x^{3 M+3}-27 (M+1)^3 (3 M+1) w_1\right)}{\left(x^{3 M+3}+27 (M+1)^3 w_1\right){}^2} \psi (x) \\
    +\frac{27 (M+1)^2 x^{6 M+3} \left(x^{3 M+3}-27 (M+1)^3 (3 M+2) w_1\right)}{\left(x^{3 M+3}+27 (M+1)^3 w_1\right){}^3} \psi (x) = 0~.
    \label{excitedODExvars}
\end{multline}
The terms in the first line are exactly those that appear in the vacuum ODE in \eqref{scalarLaxthirdorder}. From the discussion in Section \ref{vacuumODE} it follows that one has to take the adjoint of this differential equation and perform a WKB analysis to extract the eigenvalues of the conserved charges in the excited states. As in the vacuum case, we find that the results obtained for the conserved charges are the same whether we perform the WKB analysis for \eqref{excitedODExvars} or its adjoint. So we continue to work with the differential equation in \eqref{excitedODExvars}.

The additional terms in the differential equation are the simplest ($L=1$) higher spin generalization of the monster potentials proposed in \cite{Bazhanov:2003ni} to compute the eigenvalues of the quantum KdV charges in the excited states. The data of the conformal field theory can be expressed in terms of the $\ell_i$, as explained in the previous sections.  From here onwards, the path to deriving the excited state eigenvalues is precisely what has been described previously. But first, we rewrite the constraints satisfied by $a_1$ and $w_1$ in terms of the CFT parameters. The constraint on $a_1$ reads:
\begin{align} 
    a_1^2 +\frac{3M+2}{M+1}a_1+ \frac{3M+1}{(M+1)^2}-\frac{3\Delta_2}{M+1} &= 0~.
    \label{a1equation}
\end{align}
The quadratic equation has two solutions, and for each solution of the parameter $a_1$, the parameter $w_1$ is given by the equation
\begin{equation}
w_1 =-\frac{\Delta _3}{M (M+1)^{3/2}} + \frac{\Delta_2}{9M(M+1)}\left(2a_1+\frac{3M+2}{M+1}\right)~.
\end{equation}
Now, at each order in the WKB expansion, the period integral of the wavefunction at that order encodes the eigenvalues of the conserved charge in the excited states. The period integrals are obtained in terms of the parameters $(a_1, w_1)$. Substituting the two independent solutions for these, we obtain the two eigenvalues in terms of the CFT parameters. These correspond to the eigenvalues of the eigenstates (of the conserved charges) that are  particular linear combinations of the states $\{L_{-1}|\Delta_2,\Delta_3\rangle, W_{-1}|\Delta_2,\Delta_3\rangle \}$. As we shall see, while the individual eigenvalues are complicated functions of the CFT data, the sum of the eigenvalues are polynomials in $(c, \Delta_2, \Delta_3)$ and will precisely match the $O(q)$ coefficient of the thermal one-point function of the conserved charge. 

The map from the ODE written in the $x$-coordinates to the $t$-coordinates most suited to the WKB analysis is the same as the one that worked for the vacuum ODE:
\be 
x= E^{-\frac{r_3}{3M}}\,t~,\qquad \epsilon= E^{-\frac{M+1}{3M}}~. 
\ee
The final version of the ODE takes the following form:
\begin{align}
 \epsilon^3\Big( y^{(3)}(t) &+\frac{r_1r_2}{r_3^2} 
 \frac{ y'(t)}{t^2}\Big)
 +\frac{1}{r_3^3} t^{-\frac{3}{r_3}(r_3+M+1) } (1-t^{\frac{3M}{r_3}} )\, y(t) \nonumber\\
&-\frac{9 (M+1) \epsilon ^3 }{r_3^2 t^2} \Bigg \{ \frac{1-27 (M+1)^3 (3 M+2) w_1 \epsilon ^3 t^{\frac{3 (M+1)}{r_3}}}{\left(1+27 (M+1)^3 w_1 \epsilon ^3 t^{\frac{3 (M+1)}{r_3}}\right)^2} \ y'(t)  \nonumber \\
&+\frac{ a_1 (M+1)  \left(2-27 (M+1)^3 (3 M+1) w_1 \epsilon ^3 t^{\frac{3 (M+1)}{r_3}}\right)}{r_3 t \left(1+27 (M+1)^3  w_1 \epsilon ^3 t^{\frac{3 (M+1)}{r_3}}\right)^2} y(t) \nonumber \\ 
&+ \frac{27 (M+1)^3 w_1 \epsilon ^3 (3 M (r_3-3 M-4)+r_3-5) t^{\frac{3 (M+1)}{r_3}}+3 M - r_3 + 2}{ r_3 t \left(1+27 (M+1)^3 w_1 \epsilon ^3 t^{\frac{3 (M+1)}{r_3}}\right)^3} y(t)\nonumber\\
&+\frac{729 (M+1)^6 (3 M+2) (r_3+1) w_1^2 \epsilon ^6 t^{\frac{6 (M+1)}{r_3}} }{ r_3 t \left(1+27 (M+1)^3 w_1 \epsilon ^3 t^{\frac{3 (M+1)}{r_3}}\right)^3} y(t)
 \Bigg \}= 0~.
\end{align}
The WKB analysis proceeds exactly as before. We plug in the  exponential ansatz: 
\begin{equation}
   y(t) = \exp \bigg [\int^t dt \ \sum_{i \geq 0}\epsilon^{i-1} a_i(t) \bigg]
\end{equation}
into the differential equation. We collect the terms at each order in $\epsilon$, and solve for the $a_n(t)$ recursively. The period integrals of the $a_n(t)$ encode the eigenvalues of the conserved charges in the $L=1$ excited states, up to some constant coefficients:
\begin{equation}
    \widehat{I}_{n-1}=  \int_0^1\, dt\, a_n(t)~.
\end{equation} 

\subsection{Eigenvalues and Eigenstates at Level One}\label{excited_eigen}

The first non-trivial eigenvalue is given by
\be 
I_1^{(L=1)} = \Delta_2 - \frac{c}{24} + 1~,
\ee 
which is consistent with the fact that the conformal dimension of the excited state is one more than that of the primary in the module. The next charge is $\mathbf{I}_2 \equiv W_0$, whose action on the level one subspace we compute explicitly:
\be
\begin{aligned}
    \mathbf{I}_2 L_{-1}\rvert \Delta_2,\Delta_3 \rangle &= \Delta_3 L_{-1} \rvert \Delta_2,\Delta_3 \rangle + 2 W_{-1} \rvert \Delta_2,\Delta_3 \rangle ~,\\ 
    \mathbf{I}_2 W_{-1}\rvert \Delta_2,\Delta_3 \rangle &= \frac{1}{15}\left(2-\frac{1}{b^2}+10 \Delta_2\right) L_{-1} \rvert \Delta_2,\Delta_3 \rangle + \Delta_3 W_{-1} \rvert \Delta_2,\Delta_3 \rangle ~.
\end{aligned}
    \label{I2onbasis}
\ee 
Since the charges are mutually commuting, we can use this result to compute the level one eigenbasis for the Boussinesq charges:
\begin{equation}
\begin{aligned}
    \rvert e_1 \rangle &=  -\frac{\sqrt{-c+32 \Delta _2+2}}{4 \sqrt{6}} L_{-1}\rvert \Delta_2,\Delta_3 \rangle + W_{-1} \rvert \Delta_2,\Delta_3 \rangle~, \\  \rvert e_2 \rangle &=  \frac{\sqrt{-c+32 \Delta _2+2}}{4 \sqrt{6}} L_{-1}\rvert \Delta_2,\Delta_3 \rangle + W_{-1} \rvert \Delta_2,\Delta_3 \rangle~.
\end{aligned}
\label{leveloneEVs}
\end{equation}
We introduce the notation
\begin{equation}
    \mathbf{I}_n \rvert e_i \rangle = I^{(1)}_{n,i}  \rvert e_i \rangle~,
\end{equation}
where the index $i$ distinguishes the two eigenvectors at the first excited level. For $n=2$, the two eigenvalues are easily calculated  from \eqref{I2onbasis} to be
\begin{align}
    I_{2,i}^{(1)} = \Delta_3 \pm \frac{\sqrt{-c+32 \Delta _2+2}}{2 \sqrt{6}}~.
\end{align}
On the other hand, the excited state energies calculated from the ODE/IM correspondence give us:
\begin{align}
    I_{2,i}^{(1)} = \Delta_3  -\frac{2\sqrt{M+1}}{3} \left(a_{1,i} +\frac{3 M+2}{2(M+1)}\right)~.
\end{align}
Substituting the two values of $a_{1,i}$ obtained by solving \eqref{a1equation}, and recalling the relation \eqref{cvsM} between the central charge and $M$
\be 
c = 2-\frac{24M^2}{M+1}~,
\ee 
we find a precise match with the results obtained by directly diagonalizing the operator $W_0$ on the level one subspace of the ${\cal W}_3$ module. Through this, we identify the common eigenvector of the Boussinesq charges (at level 1), whose corresponding eigenvalue is obtained by setting the parameter $a_{1}$ to each solution of \eqref{a1equation}. By summing over both eigenvalues, we obtain 
\be 
\Tr_{L=1}\mathbf{I}_2 = 2\Delta_3~.
\ee 
A similar analysis can be carried out for the higher charges, and we obtain the following expressions for the sum of the eigenvalues from the excited state ODE: 
\begin{align}
\Tr_{L=1}\mathbf{I}_4 &= 2\Delta_3\left( \Delta_2 -\frac{c-90}{24}\right)\label{tracelistI4}  ~,\\
\Tr_{L=1}\mathbf{I}_5 &=2 \Delta _2^3+18 \Delta _3^2+\frac{1}{4} (112-c) \Delta _2^2+ \frac{1}{96} \left(c^2-103 c+2670\right) \Delta _2\nonumber \\ 
&\hspace{4cm}+\frac{-7 c^3+313 c^2-40506 c+87696}{48384}\label{tracelistI5} ~,\\
\Tr_{L=1}\mathbf{I}_7 &=2\Delta _2^4+ 36\Delta_3^2\Delta_2 -\frac{1}{6} (c-156) \left(2 \Delta _2^3+9 \Delta _3^2\right)\label{tracelistI7}\\ 
&+ \frac{1}{240} \left(5 c^2-713 c+25794\right) \Delta _2^2
+\frac{\left(-10 c^3+853 c^2-93636 c+534996\right) \Delta _2}{17280}\nonumber \\
&\hspace{3cm}+\frac{10 c^4-373 c^3+107916 c^2-1475460 c+2671488}{1658880}\nonumber ~,\\
\Tr_{L=1}\mathbf{I}_8 &= 2 \Delta _3 \Delta _2^3+6 \Delta _3^3+\frac{1}{4} (178-c) \Delta _3 \Delta _2^2+\frac{1}{480} \left(5 c^2-812 c+33540\right) \Delta _3 \Delta _2\nonumber\\
&\hspace{3cm}+\frac{\left(-35 c^3+2378 c^2-521028 c+637560\right) \Delta _3}{241920}~. \label{tracelistI8}
\end{align}

\subsection{Consistency with Thermal One-Point Functions}

It is now a straightforward exercise to check that the thermal one-point functions of the conserved charges computed in the previous section are consistent with the excited state eigenvalues. The checks for ${\bf I}_1$ and ${\bf I}_2$ are trivial, so we discuss the first non-trivial case:
\be 
\label{I4thermalvev}
\langle {\bf I}_4\rangle = \Delta_3 \Big(\partial \chi(\tau) - \frac14 E_{2}(\tau) \chi\Big)~. 
\ee 
The character of the higher spin module is given by (see equation \eqref{eq:restrictedV module})
\be 
\chi(\tau) = \frac{q^{\Delta_2-\frac{c}{24}}}{\prod_n(1-q^n)^2}~.
\ee 
Expanding the thermal one-point function in \eqref{I4thermalvev} as  a power series in $q$, we obtain  
\begin{align}
    \langle {\bf I}_4\rangle &= q^{\Delta_2 - \frac{c}{24}}\left\{ \Delta_3\Big(\Delta_2 - \frac{c}{24}\Big) - \frac{\Delta_3}{4} +  2q\Delta_3\Big( \Delta_2 -\frac{c-90}{24}\Big)  +\ldots \right\}~.
\end{align}
The zeroth order term matches with the vacuum eigenvalue, while the sub-leading coefficient matches the sum of eigenvalues at the first excited level in equation \eqref{tracelistI4}. We have checked that a similar match holds for all conserved charges up to ${\bf I}_8$. For ${\bf I}_{10}$ we have checked this for the special case of $M=1$ (which corresponds to $c=-10$), but with arbitrary $\Delta_2$ and $\Delta_3$.  These are important consistency checks (at the sub-leading order) of the thermal one-point functions in the previous section.

\section{Fixing the Higher Boussinesq Currents } \label{two_point}

In this section, we shall show how the form of the quantum current $J_8$ can be fixed by using the excited state eigenvalues and two-point thermal correlators. We begin by recalling the form of the current $J_8$:
\begin{multline} \label{J8}
J_8 = (T(T(TT))) +18 (T(WW)) +\frac{3}{16}(5c+46)(2\pi)^2  (W'W') \\
-\frac{1}{7680}(5c^2-28c +1124) (2\pi)^4 (T''T'')\\ 
+ (2\pi)^2\, \alpha\, \left( (T(T'T')) - 3 (W'W') + (2\pi)^2 \frac{c-34}{96}(T''T'')\right)~,
\end{multline}
where $\alpha$ is a central charge dependent constant which cannot be determined by the thermal one-point function $\langle \mathbf{I}_7 \rangle$. We know that this coefficient has to be non-zero since we have computed the $c$-dependent component of $\alpha$ in \eqref{classical_limit}. The combination of fields in the last line of \eqref{J8} does not contribute to the trace but does have non-trivial matrix elements on the level one subspace of the ${\cal W}_3$ module. Therefore, we may determine $\alpha$ by computing a trace weighted by an operator that does not act identically on the basis states of the level one subspace. The simplest such operator is the KdV charge 
\be 
\mathbf{Q}_3 = \int_0^1~ du~ (TT)(u)~,
\ee
whose operator form is given by: 
\begin{equation}
    \mathbf{Q}_3 \equiv ((TT)(u))_0 = L_{0}^{2}-\frac{c+2}{12}L_{0}+\frac{c(5c+22)}{2880}+2\sum_{n\ge0}L_{-n}L_{n}~.  
\end{equation}
We now consider the trace involving the KdV charge and any of the Boussinesq charges at the first excited level: 
\begin{align}\label{TT_inserted}
\Tr_{{\cal H}^{(1)}} \Big( \mathbf{I}_n \mathbf{Q}_3 \Big) ~.
\end{align}
We compute the trace in two distinct ways: directly using the operator formalism, and more indirectly, via calculating the thermal correlators involving the various terms in $J_9(u_1)$ and $(TT)(u_2)$, and by extracting the subleading coefficient in the $q$-expansion, after appropriate normal ordering. 

We work with the operator method first. At the first excited level, one can compute the action of ${\bf Q}_3$ on the basis states: 
\begin{align}
    \mathbf{Q}_3 L_{-1}\rvert \Delta_2,\Delta_3 \rangle &= \left(\frac{5 c^2-218 c+2400}{2880}-\frac{1}{12} (c-70) \Delta _2+\Delta _2^2 \right) L_{-1}\rvert \Delta_2,\Delta_3 \rangle~, \\ 
    \mathbf{Q}_3 W_{-1}\rvert \Delta_2,\Delta_3 \rangle  &=  6 \Delta_3 L_{-1}\rvert \Delta_2,\Delta_3 \rangle  \nonumber\\ 
    & \hspace{1cm}+\left( \frac{5 c^2-218 c+2400}{2880}-\frac{1}{12} (c-22) \Delta _2+\Delta _2^2 \right) W_{-1}\rvert \Delta_2,\Delta_3 \rangle ~.
\end{align}
To calculate the action of the conserved charges on the basis, we recall the discussion from Section \ref{excited_eigen} and the derivation of the eigenstates $|e_i\rangle$ that simultaneously diagonalize all the Boussinesq charges in \eqref{leveloneEVs}. It is straightforward to derive the following relations:
\begin{align}
    \mathbf{I}_n L_{-1}\rvert \Delta_2,\Delta_3 \rangle &= \frac{( I_{n,1}^{(1)}+I_{n,2}^{(1)})}{2} L_{-1}\rvert \Delta_2,\Delta_3 \rangle +   \frac{2 \sqrt{6}~(I_{n,2}^{(1)} -I_{n,1}^{(1)} )}{\sqrt{-c+32 \Delta _2+2}}  W_{-1}\rvert \Delta_2,\Delta_3 \rangle~, \\  \mathbf{I}_n W_{-1}\rvert \Delta_2,\Delta_3 \rangle &=  \frac{\sqrt{-c+32 \Delta _2+2}}{8 \sqrt{6}}~(I_{n,2}^{(1)} -I_{n,1}^{(1)} ) L_{-1}\rvert \Delta_2,\Delta_3 \rangle+\frac{( I_{n,1}^{(1)}+I_{n,2}^{(1)})}{2} W_{-1}\rvert \Delta_2,\Delta_3 \rangle ~.
\end{align}
The trace of the product of these two charges can now be computed for all $n$ and we obtain 
\begin{multline}
    \Tr_{{\cal H}^{(1)}} \Big( \mathbf{I}_n \mathbf{Q}_3 \Big)  = \left(\Delta _2^2 +\frac{\left(5 c^2-218 c+2400\right) }{2880}-\frac{1}{12} (c-46) \Delta _2\right)( I_{n,1}^{(1)}+I_{n,2}^{(1)})\\ +\frac{12 \sqrt{6} \Delta _3 }{\sqrt{-c+32 \Delta _2+2}}(I_{n,2}^{(1)} -I_{n,1}^{(1)} )~.
\label{odeimi7q3}
\end{multline} 
We note that this result does not require knowing the form of the current density $J_8$ but instead relies on our knowledge of the excited state level-one eigenvalues and eigenvectors of the Boussinesq charges from the ODE/IM correspondence. Thus, the expression in \eqref{odeimi7q3} can be thought of as the ODE/IM prediction for the level-one contribution to the trace of the product of charges in the first excited level.  

We now independently calculate the two-point correlator, but now using the explicit form of the current density $J_8$ in \eqref{J8}. This is done by normal ordering appropriately the required $n$-point correlators of the stress tensor and the spin 3 field, which in turn are computed via the Zhu recursion formula. For details, we refer the reader to Appendix \ref{Apen:Twopint}.  By comparing the result of this computation for the ansatz current given in  \eqref{J8} and \eqref{odeimi7q3}, we find a perfect match with the ODE/IM result if we fix $\alpha$ to be
\begin{equation}
    \alpha = \frac{1}{16}\left(3c-14\right)~.
\end{equation}
We note that this is consistent with the linear-in-$c$ piece deduced from the classical limit of the current in section \ref{classical_limit} (see the expression for $\alpha$ in \eqref{largeclimitofcoefficients}). The analysis in this section fixes $\alpha^{(0)} = - \frac{7}{8}$. In conclusion, we have completely determined the Boussinesq current $J_8$ (up to total derivatives), and it is given by:
\begin{multline}
     J_8 = (T(T(TT))) + 18 (T(WW)) + \frac{5 c^2-276 c+628}{3840}\,(2\pi)^4\, (T''T'') \\ +\frac{(3 c-14)}{16}\, (2\pi)^2\, (T(T'T')) +\frac{3}{8}\, (2\pi)^2\, (c+30) (W'W') ~.
     \label{J8final} 
\end{multline}

A similar analysis for the next current $J_9$ does not lead to a solution for the coefficient $\gamma$ because it turns out that the thermal two-point function of ${\bf Q}_3$ and the operator multiplying $\gamma$ gives a vanishing result at all orders in the $q$-expansion. A similar result holds for the operators multiplying $\delta_i$ in $J_{11}$. 
However, we have checked that if we instead insert the operator $\oint (T'T')$, then it does indeed give a non-vanishing contribution, but at sub-subleading order. Thus, one would have to calculate the eigenvalues of the conserved charges in the excited states at level $L=2$, which we leave for future work. What should be evident from the analysis in this section is that, by judiciously combining the results for the thermal correlators and the eigenvalues in higher excited levels, it is possible to determine the current densities systematically.

\section{Conclusions}

The main result of this work is the calculation of the thermal one-point functions of all conserved charges up to weight ten, and  
the derivation of the higher conserved current $J_8$ in the quantum Boussinesq hierarchy. The thermal one-point functions are obtained by first calculating the thermal correlators using the Zhu recursion relations, followed by conformal normal ordering. Independently, we have also computed the eigenvalues of the conserved charges in the vacuum and first excited state of a higher spin module through the ODE/IM correspondence, and shown the mutual consistency of these two sets of results. All thermal one-point functions we have obtained are quasi-modular linear differential operators acting on the character of the higher spin module.

It is interesting to note that the information content coming from the ODE/IM correspondence is in some sense complementary to that coming from the thermal one-point functions. From the ODE/IM side, we computed the eigenvalues of the conserved charges in the vacuum and first excited levels of the higher spin module. Each of these calculations involved a WKB analysis of ordinary differential equations. As one moves on to higher excited states, the calculations get progressively more difficult, and explicit expressions in terms of the conformal field theory data, namely the central charge, conformal dimension and higher spin of the highest weight state, would be hard to obtain (since this rewriting would involve solving algebraic equations of high degree). In contrast, the thermal one-point functions contain information about all the excited levels, but only about the sum of the eigenvalues at each level. These, in turn, have simple expressions that are polynomial in the conformal field theory data. 

One curious feature of our analysis is the fact that there are trace-free combinations of normal ordered composite operators; that is, linear combinations of composite operators of a given conformal weight, whose one-point functions give a vanishing contribution to the trace. It would be important to better understand this feature of the one-point functions. As a consequence, the thermal one-point functions alone do not lead to a unique determination of the current densities. However, the excited state eigenvalues of the charges that we compute once again using the ODE/IM correspondence, along with the higher point thermal correlators come to the rescue in this case. By computing thermal two-point functions and extracting the subleading terms in the low-temperature limit, we were able to uniquely fix the current density $J_8$, whose integral gives rise to the quantum integral of motion ${\bf I}_7$ of the Boussinesq hierarchy. 
We believe that this method, which involves both the excited state eigenvalues and the thermal higher point functions, provides a systematic way to compute the higher conserved charges unambiguously. In this way we hope to fix the undetermined constants appearing in the higher currents $J_9$ and $J_{11}$ by combining the excited state eigenvalues at the next level $(L=2)$, and the thermal two-point functions.

It is important to note that our methods would only fix the form of the current densities. In order to obtain the operator that corresponds to the conserved charge, one would have to work out the zero modes of the composite operators appearing in the currents. This can be done by following the methods of   \cite{Dymarsky:2019iny}. 
Once the charge densities and conserved charges are determined, there are many interesting directions to explore. 
The calculation of higher point functions of the conserved charges is an important next step. These will be crucial to the study of the statistics of the conserved charges (see  \cite{Maloney:2018yrz} for work in the quantum KdV case). In accordance with the generalized eigenstate thermal hypothesis (ETH) for conformal field theories \cite{Lashkari:2016vgj, Dymarsky:2019etq}, in the high-temperature limit, correlators receive their dominant contributions from the states at higher levels. We expect that the higher point correlators of the conserved charges, akin to those of \cite{Maloney:2018yrz}, should factor into a product of one-point functions in the large temperature limit. 

One could also aim to obtain exact results for the generalized partition function in the large-$c$ limit for the higher spin conformal field theory \cite{Dymarsky:2018lhf, Dymarsky:2018iwx}. We have obtained the eigen functions and eigenvalues of the conserved charges at the first excited level. It would be a difficult but worthwhile exercise to attempt a similar calculation at the higher excited levels. 

Finally, we would like to recall that, in this work, we have expressed the thermal one-point functions of the conserved charges as quasi-modular differential operators acting on the character of the higher spin module, with insertions of powers of the zero mode $W_0$. These are known in closed form only for $n=1,2$ \cite{Iles:2013jha, Iles:2014gra}. Calculating these for higher values of $n$ remains an outstanding problem. These are all questions we hope to address in the future.

\section*{Acknowledgements}

We are indebted to Dileep P. Jatkar for many helpful and insightful discussions and for his help in formulating the problem analyzed in this work. We would like to thank Benjamin Doyon, Subramanya Hegde, Alok Laddha, Roji Pius, Nemani V. Suryanarayana and Jan Troost for helpful discussions. We are grateful to Renjan Rajan John for a careful reading of this manuscript and for his helpful suggestions. 
SA would like to thank the Universit\`a di Torino, Italy for their hospitality during the completion of this work. 
R.T. received partial support from the INFN project SFT and the PRIN Project No. 2022ABPBEY, with the title ``Understanding quantum field theory through its deformations''.

\appendix
\section{Quasimodular Forms and Elliptic Functions}
\label{Appen:Weiers}
We denote the Eisenstein series as $E_{2k}(\tau)$. They are defined as \cite{Handbooks}
\be
\label{eq:Eisen}
E_{2k}(\tau)=1+\frac{2}{\zeta(1-2k)}\sum_{n\geq 1}\frac{n^{2k-1}q^n}{(1-q^n)^2}~~~~\text{ with} ~~~n,k\in \mathbb{Z}^{+}~,
\ee
where $q=e^{2\pi i \tau}$. For the first few Eisenstein series, we present the power series expansions:
\begin{align}
    E_{2}(\tau) &= 1 -24q -72 q^2 -96 q^3+\ldots\\
    E_4(\tau) &= 1 + 240 q + 2160 q^2 + 6720 q^3+\ldots\\
    E_6(\tau) &= 1 - 504 q -16632 q^2-122976 q^3+\ldots
\end{align}
All $E_{2k}(\tau)$ for $k\ge 2$ transform as a weight $2k$ modular form under modular transformations, while $E_{2}$ is a quasi-modular form: 
\be
\begin{aligned}
    E_{2}\!\left(\frac{a \tau + b}{c \tau + d}\right) &= {\left(c \tau + d\right)}^{2} E_{2}\!\left(\tau\right) - \frac{6 i}{\pi} c \left(c \tau + d\right)~.\\
    E_{2 k}\!\left(\frac{a \tau + b}{c \tau + d}\right) &= {\left(c \tau + d\right)}^{2 k} E_{2 k}\!\left(\tau\right)~.
\end{aligned}
\ee
where $\begin{pmatrix} a & b \\ c & d \end{pmatrix} \in \operatorname{SL}_2(\mathbb{Z})$. To compute the derivatives of Eisenstein series, we set $\partial = q\frac{q}{dq}$ and recall Ramanujan identities: 
\be
\begin{aligned}
    \label{eq:ramin}
    \partial E_{2}(\tau) &=  \frac{1}{12}(E_{2}^{2}\!\left(\tau\right) - E_{4}\!\left(\tau\right))~,\\ 
    \partial E_{4}(\tau) &=\frac{1}{3}\left( E_{2}\!\left(\tau\right) E_{4}\!\left(\tau\right) - E_{6}\!\left(\tau\right)\right)~,\\
    \partial E_{6}(\tau) &=\frac12\left(E_{2}\!\left(\tau\right) E_{6}\!\left(\tau\right) - E_{4}^{2}\!\left(\tau\right)\right)~.
\end{aligned}
\ee
The Weierstrass functions are defined for $k \geq 1$, as\cite{zhu:1990} 
\be
\label{eq:weiersP}
\mathcal{P}_k(x,q)=\frac{(2\pi i)^k}{(k-1)!}\sum_{n\neq 0}\frac{n^{k-1}x^n}{(1-q^n)}~.
\ee
 These expansions converge for $|q| < |x| < 1$. We write $x=e^{2\pi i u}$, where $u$ is the coordinate on a cylinder with a period $1$. The derivative with respect to $u$ is given by
\be
\label{eq:weiersPD}
\partial_u\mathcal{P}_k(e^{2\pi i u},q)=k\mathcal{P}_{k+1}(e^{2\pi i u},q)~.
\ee
The relationship between $\mathcal{P}(e^{2\pi i u},q)$ and Weierstrass functions $\rho_{k}(u,\tau)$ are given by \cite{zhu:1990}:
\be
\begin{aligned}
\label{eq:weiersPrho}
\mathcal{P}_1(e^{2\pi i u},q)&=-\rho_1(u,\tau)+2\zeta(2)E_{2}(\tau)u-i\pi ~,\\
\mathcal{P}_2(e^{2\pi i u},q)&=\rho_2(u,\tau)+2\zeta(2)E_{2}(\tau)~,\\
\mathcal{P}_k(e^{2\pi i u},q)&=
(-1)^k\rho_k(u,\tau)~~~\text{for}~k\geq2 ~,
\end{aligned}
\ee
where $q=e^{2\pi i\tau}$ and $E_{2k}(\tau)$ is the Eisenstein series. The Laurent expansion of $\rho_{k}(u,\tau)$ near small $u$ is given by 
\be
\label{eq:weiersrho}
\rho_k(u,\tau)=\frac{1}{u^k}+(-1)^{k}\sum_{n=1}^{\infty}\binom {2n+1}{ k-1}2\zeta(2n+2)E_{2n+2}(\tau)u^{2n+2-k}~,
\ee
with $\zeta(n)$ the Rienmann zeta function. Note that the above sum vanishes for $2n+2-k<0$. 

\section{Thermal Correlators from Recursion}
\label{Apen:Zhu}

\subsection{Vertex Operators and Square Modes}

\noindent The vertex operator associated to a CFT state $a$ is a local operator whose action on the CFT vacuum $\Omega$ is given by:
\begin{equation}
    V(a,z) \Omega = e^{L_{-1}z}a ~.
\end{equation} 
For a system of local vertex operators, the action on the vacuum given above determines the vertex operator completely \cite{Goddard:1989dp} (see also \cite{Gaberdiel:1999mc}). A formal series expansion of the operator associated with a state $a$ in terms of the coordinate on the plane $z$ defines the plane modes $a_n$ for it:
 \be
V(a,z)=\sum_{n\in\mathbb{Z}}\frac{a_{n}}{z^{n+h_{a}}}~,
\ee
where $h_a$ is the conformal dimension of the field $V(a,z)$. Since we compute the correlator of fields on the torus, it would be natural to use vertex operators and mode expansions in coordinates that have the right periodicity properties built in. Therefore, we transform vertex operators to the cylinder through a conformal map $z=e^{2\pi i u}$, $u$ being the coordinate on a cylinder of circumference $1$. The conformal map from the plane to the cylinder has the following effect on a primary operator $V(a,z)$ : 

\begin{equation}
    V(a,z) \rightarrow  V(e^{2\pi i u L_0} a,e^{2\pi i u})~.
\end{equation}
We now introduce a new type of vertex operator $V[a,u]$ on the cylinder through the following transformation:
\begin{equation}
    V[a,u]=V(e^{2 \pi i u L_0 } a, e^{2 \pi i u}-1)~.
\end{equation}
This type of vertex operator naturally appears in the computation of the operator product expansions of vertex operators on a cylinder\cite{tuite,Mason:2009xt}. They play an important role in the Zhu recursion relations, which relate n-point correlators on the torus to lower-point correlators. A formal series expansion of $V[{a},u]$ in the cylinder coordinates defines the square modes for us:
\begin{equation}
    V[{a},u]=\sum_{n\in\mathbb{Z}}\frac{a[n]}{u^{n+h_{a}}}~.
\end{equation}
Explicitly, the square bracket modes can be written in terms of plane modes as:
 \be
 \label{eq:squaremodes1}
 a[n]=\frac{1}{(2\pi i)^{n+1}}\sum_{j\ge n+1-h_a}c(h_{a},j+h_{a}-1,n)a_{j}~,
 \ee
where the expansion coefficients are given by
  \be
  \label{eq:coeff}
  (\log(1+z))^{s}(1+z)^{h-1}=\sum_{j\geq s}c(h,j,s)z^{j}~.
  \ee
This formula will be used for instance to compute the square modes of $W(z)$, whose vertex operator representation is given by  $V(W_{-3}\Omega, z)$. Here we tabulate just the first few of these square modes :
\be
\begin{aligned}
\label{eq:WsquareModes}
W[0]&=(2\pi i)^{-1}\Big(W_{-2}+2W_{-1}+W_{0}\Big) ~,\\
W[1]&=(2\pi i)^{-2}\Big( W_{-1}+\frac{3}{2} W_0+\frac{1}{3} W_1-\frac{1}{12} W_2+\frac{1}{30} W_3-\frac{1}{60}W_4+\ldots\Big)~,\\
W[2]&=(2\pi i)^{-3}\Big(W_{0}+W_{1}-\frac{1}{12}W_{2}+\frac{1}{90}W_{4}+\ldots\Big) ~,\\
W[3]&=(2\pi i)^{-4}\Big(W_{1}+\frac{1}{2}W_{2}-\frac{1}{4}W_{3}+\frac{1}{8}W_{4}+\ldots\Big) ~.\\
\end{aligned}
\ee
In contrast to primary fields, general vertex operators transform non-trivially under conformal transformations \cite{Gaberdiel:1994fs}. For instance, under the aforementioned map from the plane to the cylinder, the energy-momentum tensor transforms as:
\begin{equation} \label{eq:StressTensor_Cyl}
 V(L_{-2}\Omega,z)\rightarrow  (2\pi i)^2 z^2 V(L_{-2}\Omega,z) - (2\pi i)^2 V\left(\frac{c}{24}\Omega,z\right)~.
\end{equation}
This motivates the introduction of the  state $\tilde\omega$ and its modes defined through 
\begin{equation}
    \tilde\omega := \left(L_{-2} -\frac{c}{24} \right)\Omega \ , \quad \text{with} \quad V(\tilde\omega,z) = \sum_{n} \tilde\omega_n z^{-n-1} = \sum_{n} L_n z^{-n-2} - \frac{c}{24}~.
\end{equation}
The conformal dimension of the quasi-primary $ \tilde\omega$ is $2$, and the square modes of the stress tensor on the torus are defined through the formal series expansion in $u$ of the following vertex operator:
\begin{eqnarray}
   V(e^{2\pi i u L_0}  \tilde\omega,e^{2\pi i u}-1) ~ .
\end{eqnarray}
Using the power series expansion discussed previously, the square modes take the form:
\begin{equation}
    L_{[-n]} := (2\pi i)^2 \tilde\omega[n+1] = (2\pi i)^{-n} \sum_{j \ge n+1} c(2,j,n+1)L_{j-1} - (2\pi i)^2 \frac{c}{24} \delta_{n,-2}~.
\end{equation}
Here we have defined the modes $L_{[-n]}$ that satisfy the Virasoro algebra. Evaluating these square modes for the first few values of $n$, we get:
\be
\begin{aligned}
\label{eq:LsquareModes}
L_{[-1]}&=(2\pi i)^2\tilde{\omega}[0]=(2\pi i)(L_{-1}+L_{0}) ~,\\
L_{[0]}&=(2\pi i)^2\tilde{\omega}[1]=(L_{0}+\frac{1}{2}L_{1}-\frac{1}{6}L_{2}+\frac{1}{12}L_{3}-\frac{1}{20}L_4+\ldots)~,\\
L_{[1]}&=(2\pi i)^2\tilde{\omega}[2]=\frac{1}{2\pi i}(L_{1}-\frac{1}{12}L_{3}+\frac{1}{12}L_{4}+\ldots)~,\\
L_{[2]}&=(2\pi i)^2\tilde{\omega}[3]=\frac{1}{(2\pi i)^{2}}(L_{2}-\frac{1}{2}L_{3}+\frac{1}{4}L_{4}+\ldots)~. 
\end{aligned}
\ee

\subsection{Zhu Recursion Relation}

We will follow the notations and conventions of \cite{zhu:1996, Gaberdiel:2012yb}. The $n$-point correlation functions are denoted  
 \begin{equation}
   F((a^{1},z_1),...,(a^{n},z_{n});\tau) := 
\Tr[ V(e^{2\pi i u_1 L_0}\tilde{a}_1,e^{2\pi i u_1})... V(e^{2\pi i u_n L_0}\tilde{a}_n,e^{2\pi i u_n})q^{L_{0}-\frac{c}{24}}]~.
 \end{equation}
Here, the $z_i:= e^{2\pi i u_i}$ are coordinates on the plane. For later purposes, it is more convenient to write the recursion for the $n$-point correlation function,  with $l$ insertions of the zero mode of another field $b$. As shown in \cite{Gaberdiel:2012yb}, this  satisfies the (modified) Zhu recursion relation: 
\be
\begin{aligned}
\label{eq:zhu}
&F(b_{0}^{l};(a^{1},z_1),...,(a^{n},z_{n});\tau)=F(b_{0}^{l}a_{0}^{1};(a^{2},z_2),...,(a^{n},z_{n});\tau)+\\&\sum_{i=0}^{l}\sum_{j=2}^{n}\sum_{m=0}^{\infty}\binom l ig^{i}_{m+1}(z_{j1}, q)F(b_{0}^{l-i};(a^{2},z_2),...,(d^{i}[m]a^{j},z_{j}),...,(a^{n},z_{n});\tau)~,
\end{aligned}
\ee
where $z_{ij}=\frac{z_i}{z_j}$ and the functions  $g_{k}^{i}$ are defined in terms of the Weierstrass functions\footnote{More details on this can be found in Appendix A of \cite{Maloney:2018hdg}.} $\mathcal{P}_{k}(x,q)$ as
\be
g^{i}_{k}(x,q)=\frac{(2\pi i)^{k}}{(k-1)!}\sum_{n\neq 0}n^{k-i-1}x^{n}\partial^{i}\frac{1}{(1-q^n)}=(2\pi i)^{2i}\frac{(k-i-1)!}{(k-1)!}\partial_i\mathcal{P}_{k-i}(x,q)~.
\ee
In the recursion, the mode denoted $d^{i}[m]$ is  defined to be 
\be
d^{i}[m]=(-1)^{i}((b[0])^{i}a^{1})[m]~.
\ee
To find the square modes of composite objects of the form $(b[n])a$~, we use the identity (see equation $(4.2.4)$ of \cite{zhu:1996})
\be
\label{eq:zhusquare}
(b[n]a)[m]=\sum_{i}\binom{n}{i}\Big((-1)^ib[n-i]a[m+i]-(-1)^{n+i}a[n+m-i]b[i]\Big)~.
\ee

\subsubsection{Thermal Expectation Values of Zero Modes}

As we show in many examples, the Zhu recursion eventually reduces a higher point correlation function to the thermal expectation values of zero modes of the various operators. So it is useful to tabulate these for the higher spin conformal field theory. The character in the higher spin module is defined to be 
\be
\chi(\tau)= \Tr \Big(q^{L_0 - \frac{c}{24}}\Big) = \frac{q^{\Delta_2-\frac{c}{24}}}{\prod_{n=1}^{\infty}(1-q^n)^2}~. 
\ee 
The zero mode of the energy-momentum tensor on the torus is given by 
\be
\tilde\omega_{(0)} := \int du \ V(e^{2\pi i u L_0}\tilde{\omega},e^{2\pi i u}) = L_0-\frac{c}{24}~.
\ee
Thus, the insertion of the zero mode is equivalent to a derivation:
\be 
\Tr \Big(\tilde\omega_{(0)}^n\, q^{L_0 - \frac{c}{24}}\Big) =(q\frac{d}{dq})^n \chi(\tau) = \partial^n \chi(\tau)~,
\ee 
where we have defined $\partial = q\frac{d}{dq}$. 

The insertion of the $W_0$ operator is more complicated and these have been evaluated for $n=1,2$ exactly, while for $n>2$, the first few terms in the power series expansion have been calculated in \cite{Iles:2013jha, Iles:2014gra}. We list the results that are of relevance to our calculations:
\begin{align}
 \Tr W_0 q^{L_0-\frac{c}{24}} &= \Delta_3 \chi(\tau)=q^{\Delta_2-\frac{c}{24}}\Big(\Delta_3+2\Delta_3q+5\Delta_3q^2+\mathcal{O}(q^3)\Big)~,\label{vevofW0}\\
\Tr W_0^2 q^{L_0-\frac{c}{24}} &= \Big(\Delta_3^2-\frac{1}{9}(\Delta_2-\frac{c-2}{24})E^{\prime}_{2}(\tau)+\frac{1}{27}E^{\prime\prime}_{2}(\tau)+\frac{1}{6}\frac{c+30}{1440}E^{\prime}_{4}(\tau)\Big)\chi(\tau) \nonumber\\
&=q^{\Delta_2-\frac{c}{24}}\Big(\Delta^2_3+(2\Delta^2_3+\frac{1}{16}(32\Delta_2-c+2))q+\mathcal{O}(q^2)\Big)\label{vevofW0sq}~,\\
\Tr W_0^3 q^{L_0-\frac{c}{24}} &= q^{\Delta_2-\frac{c}{24}}\Big(\Delta^3_3+\big(2\Delta^3_3+\frac{\Delta_3}{4}(32\Delta_2-c+2)\big)q+\mathcal{O}(q^2)\Big) ~.
\end{align}

\subsection{Thermal Correlators}

In this section, we illustrate how the Zhu recursion works in practice by computing some two and three-point correlators involving the energy-momentum tensor and the spin-3 current. The thermal correlators involving just the energy-momentum tensors have been worked out in detail in \cite{Maloney:2018hdg}, so after reviewing the simplest two-point correlator of the stress tensor, we shall mostly present the results for the new correlators involving the spin-3 field.

\subsubsection{$\langle T(u_1)T(u_2)\rangle$}

We start with the two-point function, which is $\langle T(u_1)T(u_2)\rangle$. For this purpose we can put  $l=0$ and $a^{1}=a^{2}=\tilde{\omega}$ in \eqref{eq:zhu} and we obtain
    \be
\begin{aligned}
\label{eq:TT1}
F((\tilde{\omega},z_1),(\tilde{\omega},z_{2});\tau)&=F(\tilde{\omega}_{(0)};(\tilde{\omega},z_2);\tau)+\sum_{m=0}^{\infty} g^{0}_{m+1}(z_{21})F((d^{0}[m]\tilde{\omega},z_2);\tau)\\
&=F(\tilde{\omega}_{(0)};(\tilde{\omega},z_2);\tau)+\sum_{m=0}^{\infty} g^{0}_{m+1}(z_{21})F((\tilde{\omega}[m]\tilde{\omega},z_2);\tau)\\
&=F((\tilde{\omega}_{(0)})^2;\tau)+\sum_{m=0}^{\infty} g^{0}_{m+1}(z_{21})F((\tilde{\omega}[m]\tilde{\omega})_0;\tau)~.\\
\end{aligned}
\ee
Using the fact that the zero mode of $\tilde{\omega}$ is $L_0-\frac{c}{24}$, we find  that $F((\tilde{\omega}_{(0)})^n;\tau)=\partial^{n} \chi(\tau)$ where $\partial=q\frac{\partial}{\partial q}$ and $\chi(\tau)$ is the reduced character defined in \eqref{eq:cha}. Since $\tilde{\omega}[0]\propto L_{[-1]}$, the action of $\tilde{\omega}[0]$ on a state corresponds to the action of a derivative with respect to the cylinder co-ordinate; thus the zero mode for such a state is vanishing. Thus, the $m=0$ term in \eqref{eq:TT1} does not contribute to the correlator. To compute the contribution of the $m>0$ terms in \eqref{eq:TT1}, we use the square modes listed in \eqref{eq:LsquareModes}. By using all of the above arguments, we find that 
\begin{align}
F((\tilde{\omega},z_1),(\tilde{\omega},z_{2});\tau)
&=F((\tilde{\omega}_{(0)})^2;\tau)+\frac{2}{(2\pi i)^{2}} \mathcal{P}_{2}(z_{21})F((\tilde{\omega}_{(0)});\tau)+\frac{c}{2(2\pi i)^{4}}\mathcal{P}_{4}(z_{21})F(\tau) \nonumber\\
&=\partial^{2} \chi(\tau)+\frac{2}{(2\pi i)^{2}} \mathcal{P}_{2}(z_{21})\partial \chi(\tau)+\frac{c}{2(2\pi i)^{4}}\mathcal{P}_{4}(z_{21})\chi(\tau)~.
\label{eq:TT}
\end{align}
Next, we perform the normal ordering using the definition (see \eqref{eq:normalorderingcontour}): 
\begin{align}
\label{TTnormalorderedv0}
    \langle (TT)(u_1)\rangle=\frac{1}{2\pi i}\oint_{u_1}\frac{du_2}{u_2-u_1}\langle T(u_1)T(u_2)\rangle~.
\end{align}
In order to do so, we use expansions of the Weierstrass functions (see \eqref{eq:weiersPrho} and \eqref{eq:weiersrho})
\begin{equation}
    \begin{aligned}
        \mathcal{P}_{2}(e^{2\pi i(u_2-u_1)},q)=&\frac{1}{(u_2-u_1)^2}+2\zeta(2)E_{2}(\tau)+6\zeta(4)E_{4}(\tau)(u_2-u_1)^2+\ldots\\
\mathcal{P}_{4}((e^{2\pi i(u_2-u_1)},q)=&\frac{1}{(u_2-u_1)^4}+2\zeta(4)E_{4}(\tau)+20\zeta(6)E_{6}(\tau)(u_2-u_1)^{2}+\ldots \\
    \end{aligned}
\end{equation}
Only the third term in each of the expansions contributes to the integral over $u_2$ in \eqref{TTnormalorderedv0}, and substituting the values $\zeta(2) =\frac{\pi^2}{6}$ and $\zeta(4)=\frac{\pi^4}{90}$, we obtain 
\be
\label{TTnormalordered}
\langle(TT(u_1))\rangle=\partial^{2} \chi(\tau)-\frac{1}{6}E_{2}(\tau)\partial\chi(\tau)+\frac{c}{1440}E_{4}(\tau)\chi(\tau)~.
\ee

\subsubsection{$\langle T(u_1)W(u_2)\rangle$}
\label{app:TW}
For this correlator we set $l=0$, $a^{1}=\tilde{\omega}$ and $a^{2}=W$ in \eqref{eq:zhu} and we obtain 
\be
\begin{aligned}
\label{eq:TW1}
F((\tilde{\omega},z_1),(W,z_{2});\tau)=&F((\tilde\omega)_0(W)_0;\tau)+\sum_{m=0}^{\infty} g^{0}_{m+1}(z_{21})F((\tilde{\omega}[m]W)_0;\tau)~. 
\end{aligned}
\ee
Using the square modes of $\tilde{\omega}$ given in \eqref{eq:LsquareModes} we can show that $\tilde{\omega}[m]W=0$ for $m\geq2$. The trace of the insertion of the zero mode $W_0$ over the restricted Verma module was computed in \cite{Iles:2013jha}. We use that result to write the first term on the right-hand side of \eqref{eq:TW1} as 
\be
F((\tilde\omega)_0(W)_0;\tau)=\Delta_3 q\frac{\partial}{\partial q}\chi(\tau)~.
\ee
Combining the aforementioned arguments gives
\be
\label{eq:TW}
F((\tilde{\omega},z_1),(W,z_{2});\tau)=\Delta_3 \partial\chi(\tau)+\frac{3}{(2\pi i)^2}\mathcal{P}_2(z_{21})\Delta_3\chi(\tau)~.
\ee
Normal ordering is done as before, and we obtain
\be
\begin{aligned}
\langle (TW(u_1))\rangle=
&\Delta_3\Big( \partial\chi(\tau) -\frac{1}{4}E_2(\tau)\chi(\tau)\Big)~.
\end{aligned}
\ee
\subsubsection{$\langle T(u_1) T(u_2) T(u_3) \rangle$}
\label{app:TTT}
We now compute a thermal correlator involving three energy-momentum tensors. We set $l=0$ and $a^{1}=a^{2}=a^{3}=\tilde{\omega}$ in \eqref{eq:zhu} and we obtain
  \begin{multline}
      F((\tilde{\omega},z_1),(\tilde{\omega},z_{2}),(\tilde{\omega},z_3);\tau)=
      \partial^{3}\chi(\tau)+\frac{2}{(2\pi i)^{2}}\Big\{\mathcal{P}_{2}(z_{21})+\mathcal{P}_{2}(z_{31})+\mathcal{P}_{2}(z_{32})\Big\}\partial^{2}\chi(\tau)\\
      +\frac{1}{(2\pi i)^{4}}\Big\{4\mathcal{P}_{2}(z_{32})\Big(\mathcal{P}_{2}(z_{31})+\mathcal{P}_{2}(z_{21})\Big)+2(2\pi i)^2 \partial\mathcal{P}_{2}(z_{32})\\
      +4\mathcal{P}_{1}(z_{31})\mathcal{P}_{3}(z_{32})-4\mathcal{P}_{1}(z_{21})\mathcal{P}_{3}(z_{32}) \\\hspace{3cm} +\frac{c}{2} \Big(\mathcal{P}_{4}(z_{21})+\mathcal{P}_{4}(z_{31})+\mathcal{P}_{4}(z_{32})\Big)\Bigg\}\partial \chi(\tau)\\
      +\frac{c}{(2\pi i)^6}\Bigg\{\frac{1}{2}(2\pi i)^2
        \partial\mathcal{P}_{4}(z_{32})
        +\mathcal{P}_{2}(z_{31})\mathcal{P}_{4}(z_{32})+\mathcal{P}_{2}(z_{21})\mathcal{P}_{4}(z_{32})\\
        \hspace{2cm}+2\mathcal{P}_{1}(z_{31})\mathcal{P}_{5}(z_{32})-2\mathcal{P}_{1}(z_{21})\mathcal{P}_{5}(z_{32})\Bigg\}\chi(\tau)~.
  \end{multline}
The normal ordering for three-point function is defined as 
\be
\langle(T(TT))(u_1)\rangle= \frac{1}{(2\pi i)^2}\oint_{u_{1}}\frac{du_2}{u_2-u_1}\oint_{u_2}\frac{du_3}{u_3-u_2}\langle T(u_1) T(u_2) T(u_3) \rangle ~.
\ee
  Note that here we have coupled Weierstrass functions in the integrand. To do the integration sequentially, we use the method discussed in\cite{Maloney:2018hdg}, where we Taylor expand $\mathcal{P}(z_{ij})$ using \eqref{eq:weiersPD} about $z_{i-1}-z_{j}=0$ if $j\neq i-1$. Then we use the expansion of the Weierstrass function in terms Eisenstein series given in \eqref{eq:weiersPrho} and \eqref{eq:weiersrho} to do the integral, where again only the simple pole will contribute. 
  
  After performing the normal ordering, we get
  \begin{multline}
  \label{eq:TTTnormal}
\langle(T(TT))(u_1) \rangle=  \partial^{3}\chi(\tau)-\frac{1}{2}E_{2}(\tau)\partial^{2}\chi(\tau)\\
+\Big(\frac{1}{24}E_{2}^{2}(\tau)+\frac{1}{40}E_{4}(\tau)+\frac{c}{480}E_{4}(\tau)\Big)\partial \chi(\tau)-\frac{c}{3024}E_{6}(\tau)
\chi(\tau)~.
  \end{multline}
  We take the low-temperature limit $q\rightarrow0$ and find
  \begin{multline}
 \label{eq:TTTzero}
     \lim_{q\rightarrow 0} q^{\frac{c}{24}-\Delta_2} \langle(T(TT))\rangle = 
\Delta _2^3-\frac{1}{8} (c+4) \Delta _2^2\\+\frac{1}{960} (c+2) (5 c+32) \Delta _2-\frac{c \left(35 c^2+462 c+1504\right)}{483840}~.
 \end{multline}

\subsubsection{$\langle W(u_1)W(u_2)\rangle$}

For this case we set  $l=0$, $a^{1}=W$ and $a^{2}=W$ in \eqref{eq:zhu} and we have 
\begin{equation}
\begin{aligned}
\label{eq:WW1}
F((W,z_1),(W,z_2);\tau)
=&
~F(W^{2}_{0};\tau)+\sum_{m=0}^{\infty}g^{0}_{m+1}(z_{21})F((W[m]W)_0;\tau)\\	
=&
~F(W^{2}_{0};\tau)+g^{0}_{2}(z_{21})F((W[1]W)_0;\tau)\\&~~+\frac{2}{3b^2(2\pi i)^{4}}g^{0}_{4}(z_{21})F(\tilde{\omega}_{(0)};\tau)+\frac{c}{9b^2(2\pi i)^{6}}g^{0}_{6}(z_{21})\chi(\tau)~.
\end{aligned}
\end{equation}
We use the square modes of $W$ given in \eqref{eq:WsquareModes} to compute the zero mode of $W[m]W$. The calculation of the zero mode of $W[1]W$ is straightforward when $m\neq1$. When $m=1$, we have
\be
\label{eq:W[1]W}
W[1]W_{-3}\Omega=(2\pi i)^{-2}\Big[\Big(\frac{1}{5b^2}-\frac{2}{5}\Big)L_{-4}\Omega+\frac{1}{2b^{2}}L_{-3}\Omega+\frac{2}{9b^{2}}L_{-2}\Omega+\frac{c}{270b^2}\Omega+\frac{2}{3}L_{-2}^{2}\Omega\Big]~.
\ee
For all but the last term, the zero modes are easy to compute, as 
\be 
\Big(L_{-k}\Omega\Big)_0=(-1)^k(k-1)L_0~. 
\ee 
To compute the zero mode of $(L_{-2})^2\Omega$, we use the fact that it is identical to the zero mode of $(TT)(z)$ on the plane, which we computing using the formula for the $n$th mode of $(AB)(z)$ in terms of the modes of $A(z)$ snd $B(z)$ \cite{DiFrancesco:1997nk} :
\begin{equation}
    (AB)_{m} = \sum_{n\leq - h_A} A_n  B_{m-n} + \sum_{n>-h_A} B_{m-n}A_n
\end{equation}
We get
\be
\label{eq:zeroL_-2}
\Big(L^{2}_{-2} \, \Omega\Big)_0= (TT)_0 = L^2_{0}+2L_{0}+2\sum_{n\ge0}L_{-n}L_{n}~.
\ee
Collecting all the zero modes, and substituting into \eqref{eq:WW1}, we get
\begin{equation}
\begin{aligned}
\label{eq:WW2}
F((W,z_1),(W,z_2);\tau)
=&
~F(W^{2}_{0};\tau)+\frac{1}{(2\pi i)^{2}}\mathcal{P}_{2}(z_{21})\Big(\frac{2}{3}\partial^{2}\chi(\tau)-\frac{1}{9}\partial \chi(\tau)+\frac{c}{2160}\chi(\tau)\Big) \\
&+\frac{1}{(2\pi i)^{2}}\mathcal{P}_{2}(z_{21})\left\langle\frac{4}{3}\sum_{n\ge0}L_{-n}L_{n}\right\rangle \\
&+\frac{2}{3b^2(2\pi i)^{4}}\mathcal{P}_{4}(z_{21})\partial \chi(\tau)+\frac{c}{9b^2(2\pi i)^{6}}\mathcal{P}_{6}(z_{21})\chi(\tau)~.
\end{aligned}
\end{equation}
Finally we apply the technique described in \cite{Maloney:2018hdg} to compute $\left\langle\sum_{n\ge 0}L_{-n}L_{n}\right\rangle$ by moving $L_{-n}$ through the trace, and obtain  
\begin{equation}
    \begin{aligned}
    \label{eq:LnL-n}
        \left\langle\sum_{n\ge 0}L_{-n}L_{n}\right\rangle 
       &=\frac{1-E_{2}(\tau)}{12}\partial \chi(\tau)+c\frac{E_{4}(\tau)-1}{2880}\chi(\tau)~.
    \end{aligned}
\end{equation}
Substituting this into the correlator we have
\begin{equation}
\begin{aligned}
\label{eq:WW}
F((W,z_1),(W,z_2);\tau)
=&F(W^{2}_{0};\tau)+\frac{1}{(2\pi i)^{2}}\mathcal{P}_{2}(z_{21})\Big(\frac{2}{3}\partial^{2}\chi(\tau)-\frac{1}{9}\partial \chi(\tau)+\frac{c}{2160}\chi(\tau)\Big) \\
&+\frac{1}{(2\pi i)^{2}}\mathcal{P}_{2}(z_{21})\Big(\frac{1-E_{2}(\tau)}{12}\partial \chi(\tau)+c\frac{E_{4}(\tau)-1}{2880}\chi(\tau)\Big) \\
&+\frac{2}{3b^2(2\pi i)^{4}}\mathcal{P}_{4}(z_{21})\partial \chi(\tau)+\frac{c}{9b^2(2\pi i)^{6}}\mathcal{P}_{6}(z_{21})\chi(\tau)~.
\end{aligned}
\end{equation}
We normal order the above correlator and we get
\begin{multline}
\langle (WW(u_1))\rangle=\langle (W_{0})^{2}\rangle-\frac{1}{18}E_{2}(\tau)\partial^{2}\chi(\tau)+\frac{1}{17280}(160E^{2}_{2}+(22 + 5 c)E_{4}(\tau))\partial\chi(\tau)\\
-\frac{c}{4354560}(168E_{2}(\tau)E_{4}(\tau)+(22 + 5 c)E_{6}(\tau))\chi(\tau)~.
\end{multline}
The thermal one-point function of $W_0^2$ and its $q$-expansion has been listed previously in \eqref{vevofW0sq}. 
Substituting this into the normal ordered one-point function, and taking the low-temperature limit, we get
\be
\lim_{q \rightarrow 0} q^{\frac{c}{24}-\Delta_2}\langle(WW)(u_1)\rangle =\Delta_{3}^2-\frac{1}{18}\Delta_{2}^{2}+\Big(\frac{17 c}{3456}+\frac{91}{8640}\Big)\Delta_{2}-\frac{191 c^2}{1741824}-\frac{2101 c}{4354560} ~.
\ee

\subsubsection{$\langle T'(u_1)T'(u_2)\rangle$}
To compute the thermal correlator involving the derivatives of $T$, we us the fact that $\tilde{\omega}[0]$ operating on $T$ is $T^{\prime}$. We put $l=0$ and $a^{1}=a^{2}=\tilde{\omega}[0]\tilde{\omega}$ in \eqref{eq:zhu} and we obtain
\be
\begin{aligned}
F((\tilde{\omega}[0]\tilde{\omega},z_1),(\tilde{\omega}[0]\tilde{\omega},z_{2});\tau)=-\frac{12}{(2\pi i)^{6}} \mathcal{P}_{4}(z_{21})\partial \chi(\tau)-\frac{10c}{(2\pi i)^{8}}\mathcal{P}_{6}(z_{21})\chi(\tau) ~.   
\end{aligned}
\ee
Here we use the fact that the zero mode of $\tilde{\omega}[0]\tilde{\omega}$ is zero. We perform the normal ordering and get
\be
(2\pi)^2\langle (T^{\prime}T^{\prime})(u_1)\rangle=\frac{ E_{4}(\tau )}{60 }\partial \chi(\tau )-\frac{c E_{6}(\tau )}{3024}\chi(\tau)~.
\ee
In the low-temperature limit, we obtain 
\be
\lim_{q \rightarrow 0}~q^{\frac{c}{24}-\Delta_2}~(2\pi)^2 \langle(T'T')(u_1)\rangle=\frac{\Delta _2}{60}-\frac{31 c}{30240}~.
\ee
 
\subsection{Thermal One-Point Function of Composite Operators}

\subsubsection{Weight Eight}
\label{app:listofthermalvevsI7}
Here we present the thermal one-point functions of composite objects which are relevant to compute the thermal one-point function of $I_7$.
    \begin{align*}
        \langle(T(T(TT)))\rangle=&\partial^{4}\chi(\tau)-E_{2}(\tau)\partial^{3}\chi(\tau)+\frac{1}{240}((c+32) E_4(\tau )+60 E^2_2(\tau))\partial^{2}\chi(\tau)\\
        &-\frac{1}{4320}(3 (c+32) E_4(\tau ) E_2(\tau )+10 (c+6) E_6(\tau )+60 E^3_2(\tau ))\partial\chi(\tau)\\
        &+\frac{1}{4838400}c \left((7 c+1024) E^{2}_{4}(\tau )+10 (7 c+32) E_8(\tau )\right)\chi(\tau) ~,\\
        \langle(T(WW))\rangle=&-\frac{1}{18}E_{2}(\tau)\partial^{3}\chi(\tau)+\frac{1}{17280}\Big((5c+214) E_4(\tau )+560 E^2_2(\tau)\Big)\partial^{2}\chi(\tau)\\
        &-\frac{1}{4354560}\Big(42E_{2}(\tau)(320E^{2}_{2}(\tau)+3 (98 + 3 c)E_4(\tau ))\\&+(150 + c) (22 + 5 c) E_6(\tau )\Big)\partial\chi(\tau)\\
        &+\frac{c}{174182400} \Big(560 E^{2}_{2}(\tau)E_{4}(\tau )+4 (446 + 25 c)  E^{2}_{4}(\tau )+2240 E_{2}(\tau)E_{6}(\tau )\\&+15 (22 + 5 c)  E_8(\tau )\Big)\chi(\tau)+\Big(q\frac{\partial}{\partial q}-\frac12 E_{2}(\tau)\Big)\langle W_{0}^{2}\rangle ~,
        \\(2\pi)^2\langle(T(T^{\prime}T^{\prime}))\rangle=&\frac{1}{60}E_4(\tau )\partial^{2}\chi(\tau)-\frac{1}{15120}\Big(42E_{2}(\tau)E_{4}(\tau)+5(14+c)E_{6}(\tau)\Big)\partial\chi(\tau)\\&+\frac{c}{604800}(100E^{2}_{4}(\tau)-9E_{8}(\tau))\chi(\tau)~,\\ 
        (2\pi)^4\langle(W^{\prime}W^{\prime})\rangle=&\frac{1}{180}E_4(\tau )\partial^{2}\chi(\tau)-\frac{1}{181440}\Big(168E_{2}(\tau)E_{4}(\tau)+5 (22 + 5 c)E_{6}(\tau)\Big)\partial\chi(\tau)\\&+\frac{c}{4147200}\Big(16E^{2}_{4}(\tau)+ (22 + 5 c)E_{8}(\tau)\Big) ~,\\ (2\pi)^4\langle(T^{\prime\prime}T^{\prime\prime})\rangle=&-\frac{1}{126}E_6(\tau )\partial\chi(\tau)+\frac{c}{2880}E_{8}(\tau)\chi(\tau) ~.
\end{align*}
Using the above results, we can show 
\be
\left \langle
 (T(T'T')) - 3 (W'W') + (2\pi)^2 \frac{c-34}{96}(T''T'')
 \right\rangle = 0~.
 \ee 

\subsubsection{Weight Nine}
\label{app:listofthermalvevsI8} 
Next, we present the thermal one-point functions of composite objects which are relevant to compute the thermal one-point function of $I_8$. 
{\allowdisplaybreaks
    \begin{align*}
    \label{listofthermalvevsI8} 
        \langle(T(T(TW)))\rangle=&\partial^{3}\langle W_{0}\rangle-\frac{5}{4} E_2(\tau)\partial^{2}\langle W_{0}\rangle
        +\frac{1}{480}(200E^{2}_2(\tau )+ (c+108) E_4(\tau ))\partial\langle W_{0}\rangle\\
        -&\frac{1}{120960} \left(4200 E^{3}_{2}(\tau )+63 (108 + c)  E_2(\tau )E_4(\tau)+130 (30 + c)E_{6}(\tau)\right)\langle W_{0}\rangle ~, \\
        \langle(W(WW))\rangle=&\langle W^{3}_{0}\rangle-\frac{1}{6}E_{2}(\tau)\partial^{2}\langle W_{0}\rangle+\frac{1}{5760}(400E^{2}_{2}(\tau )+(166 + 5 c)E_{4}(\tau))\partial\langle W_{0}\rangle\\
        -&\frac{1}{1451520}(5040E^{3}_{2}(\tau )+21 (978 + 23 c)E_{2}(\tau)E_{4}(\tau)\\&+(3930 + c (217 + 5 c))E_{6}(\tau))\langle W_{0}\rangle~,\\
        (2\pi)^2 \langle(T(WT^{\prime\prime}))\rangle=&-\frac{1}{24}E_{4}(\tau)\partial\langle W_{0}\rangle+\frac{1}{6048}(63E_{2}(\tau)E_{4}(\tau)+2 (30 + c)E_{6}(\tau))\langle W_{0}\rangle ~,\\
           (2\pi)^2 \langle(W(T^{\prime}T^{\prime}))\rangle=&\frac{1}{60}E_{4}(\tau)\partial\langle W_{0}\rangle-\frac{1}{15120}(63E_{2}(\tau)E_{4}(\tau)+5 (45 + c)E_{6}(\tau))\langle W_{0}\rangle~,\\
   (2\pi)^4 \langle(T^{\prime\prime}W^{\prime\prime})\rangle=& -\frac{1}{84}E_{6}(\tau)\langle W_{0}\rangle ~.
    \end{align*}
}
Using the above results, we can show
\be
\left \langle(T(WT''))+\frac{5}{2}(W(T'T'))-\frac{1}{24}(c+55)(2\pi)^2 (T''W'') \right\rangle=0 ~.
\ee

\subsubsection{Weight Eleven}
\label{app:listofthermalvevsI10} 
Next, we present the thermal one-point functions of composite objects which are relevant to compute the thermal one-point function of $I_{10}$. 
\begin{equation*}
    \begin{aligned}
    \langle(T(T(T(TW))))\rangle=&\partial^{4}\langle W_{0}\rangle-2E_{2}(\tau)\partial^{3}\langle W_{0}\rangle+\big(\frac{1}{240} (c+152) E_4(\tau )+\frac{5}{4} E^2_2(\tau )\big)\partial^{2}\langle W_{0}\rangle\\-&
        \Big(\frac{1}{360} (c+152) E_4(\tau ) E_2(\tau )+\frac{1}{378} (2 c+81) E_6(\tau )+\frac{5}{18}  E^3_2(\tau )\Big)\partial\langle W_{0}\rangle\\+&\frac{1}{4838400}\Big((7 c^2+4882 c+175356) E^2_4(\tau )+2 (35 c^2+3439 c+36978) E_8(\tau )\\+&1680 (c+152) E_4(\tau ) E^2_2(\tau )+3200 (2 c+81) E_6(\tau )E_2(\tau )+84000 E^4_2(\tau )\Big)\langle W_{0}\rangle ~,\\
          \langle(W(W(WT)))\rangle=&\partial\langle W^{3}_{0}\rangle-\frac{3}{4} E_2(\tau)\langle W^{3}_{0}\rangle-\frac{1}{6}E_2(\tau)\partial^{3}\langle W_{0}\rangle+\frac{1}{5760}\Big((5 c+502)E_4(\tau )\\+&1040E^2_2(\tau )\Big)\partial^{2}\langle W_{0}\rangle-\frac{1}{1451520} \Big(\Big(5 c^2+1717 c+42210\Big) E_6(\tau )\\+&1008 (c+97) E_4(\tau ) E_2(\tau )+63840 E^{3}_2(\tau )\Big)\partial\langle W_{0}\rangle+\frac{1}{58060800}\times\\&\Big(10 (5 c^2+1941 c+55650) E_6(\tau ) E_2(\tau )+(235 c^2+16481 c+371826) E_4(\tau )^2\\+&(185 c^2+5989 c+51954) E_8(\tau )+280 (23 c+1926) E_4(\tau ) E_2(\tau )^2\\+&100800 E_2(\tau )^4\Big)\langle W_{0}\rangle ~,\\
        (2\pi)^6 \langle( T^{\prime\prime\prime} W^{\prime\prime\prime})\rangle=&\frac{1}{80} E_8(\tau )\langle W_{0}\rangle ~,\\
         (2\pi)^4  \langle( T^{\prime\prime\prime} (WT^{\prime}))\rangle =& \frac{1}{126} E_6(\tau )\partial\langle W_{0}\rangle-\frac{1}{20160} \Big(7 (c+9) E_8(\tau )+40 E_2(\tau )E_6(\tau )\Big)\langle W_{0}\rangle ~,\\
         (2\pi)^2\langle(W (W^{\prime}W^{\prime}))\rangle=&\frac{1}{180} E_4(\tau)\partial^{2}\langle W_{0}\rangle-\frac{1}{181440}\Big(25 (c+38) E_6(\tau )+672 E_2(\tau ) E_4(\tau )\Big)\partial\langle W_{0}\rangle\\&+\frac{1}{29030400} \Big(7 (5 c^2+299 c+6102) E_4(\tau )^2+1000 (c+38) E_6(\tau ) E_2(\tau )\\&+13440 E_4(\tau ) E_2(\tau )^2\Big)\langle W_{0}\rangle~.
         \end{aligned}
\end{equation*}
The thermal one-point functions of the remaining four composite operators can be inferred from the four trace-free relations, which we list: 
\begin{align*}
    &\left\langle\Big( (T(T(WT^{\prime\prime})))+\frac{21}{2}(W(W^{\prime}W^{\prime}))
    +\frac{(2\pi)^2}{192} (19 c+258)(T^{\prime\prime\prime}( WT^{\prime}))\right\rangle \cr
    &\hspace{9cm} +
   \left\langle (2\pi)^4\frac{(5 c^2+211 c+608)}{2880}(T^{\prime\prime\prime}W^{\prime\prime\prime})\Big)\right\rangle=0 ~,
   \end{align*}
   \begin{align*}
  \left\langle\left((T^{\prime\prime\prime}( WT^{\prime}))+(W(T^{\prime\prime}T^{\prime\prime}))-\frac{5}{6}(2\pi)^2(T^{\prime\prime\prime}W^{\prime\prime\prime})\right)\right\rangle=0~,
    \end{align*}
\begin{align*}
     \left\langle\left((T(WT^{\prime\prime\prime\prime})) +\frac{5}{2}(T^{\prime\prime\prime}( WT^{\prime}))+(2\pi)^2\frac{1}{120} (5 c-21) (T^{\prime\prime\prime}W^{\prime\prime\prime})\right)\right\rangle=0~,
    \end{align*}
    \begin{align*}
   & \left\langle \Big(T^{\prime}(T^{\prime}(TW)))
    -3(W(W^{\prime}W^{\prime})) -\frac{1}{96}(2\pi)^2(c-150)(T^{\prime\prime\prime}( WT^{\prime}))\right\rangle \cr
    &\hspace{9cm}+ 
   \left\langle (2\pi)^4\frac{(9 c-185)}{1440}(T^{\prime\prime\prime}W^{\prime\prime\prime})\Big)\right\rangle=0~.
\end{align*}

\subsection{A Two Point Correlator}
\label{Apen:Twopint}

In this section, we show how to calculate the two point correlator 
\be 
\Big\langle(TT)(u_1)J_n(v_1)\Big\rangle~,
\ee
for $n=8$ using the Zhu recursion relations. The procedure described here is fairly general and can be used to compute two-point functions of any two composite operators. 
To illustrate the ideas involved let us discuss the thermal two point function $\langle(TT)(u_1)(T(WW))(v_1)\rangle$. The other correlators involving the composite operators in $J_8$ can be computed similarly. 

The first step is to use the Zhu recursion to compute 
\be 
\Big\langle T(u_1)T(u_2)T(v_1)W(v_2)W(v_3)\big\rangle~, 
\ee 
and then perform normal ordering in a particular manner. Using the arguments used in \eqref{app:TTT}  we have that 
\begin{multline}
\langle(TT)(u_1)(T(WW))(v_1)\rangle =\frac{1}{(2\pi i)^3}\oint_{u_{1}}\frac{du_2}{u_2-u_1}\oint_{v_{1}}\frac{dv_2}{v_2-v_1}\\
\oint_{v_3}\frac{dv_3}{v_3-v_2}\langle T(u_1)T(u_2)T(v_1)W(v_2)W(v_3) \rangle . 
\end{multline}
This two-point function depends only on the difference in the positions of the insertions, and expanding it in these variables and discarding all but just the constant term gives us the zero mode. In this computation, we will encounter the product of two Weierstrass functions with the same argument, for which we are only interested in the constant piece. The constant piece is given by
\be
\mathcal{P}_{m_{1}}(x)\mathcal{P}_{m_{2}}(x)|_{x\rightarrow 0}=\frac{(2\pi i)^{m_{1}+m_{2}}(-1)^{m_{2}}}{(m_1-1)!(m_2-1)!}\zeta(3-m_1-m_2)\partial E_{m_1+m_2-2} ~.
\ee
We once again refer the reader to Appendix G of \cite{Maloney:2018hdg} for further details.

\section{Vacuum Eigenvalues of Quantum Boussinesq Charges}
\label{ListOfEigenvalues}
In this section, we list the eigenvalues of the quantum Boussinesq charges in the highest weight vector of the higher spin module labeled by $(\Delta_2, \Delta_3)$. The charges $I_{3n}$ vanish, and the rest can be related to the contour integrals in \eqref{contourintegral} by the formula
\begin{equation}
    \widehat{I}_n =c_n~ e^{-\frac{n\pi i}{3} } (1+M)^{\frac{n+1}{2}}\frac{\Gamma(\frac{n}{3M} + \frac{n}{3})\Gamma\left(-\frac{n}{3}\right) }{\Gamma(\frac{n}{3M})}~I_n~.
\end{equation}
For the first twelve charges, the $c_n$ are given by the list 
$\{1,0,3,9,0, 108\}$ for odd $n$, and the list  
$\{3,9,0,135,567,0\}$ for even $n$.
The eigenvalues are given by
\begin{subequations}
\begin{align}
I_1 & = \Delta_2 - \frac{c}{24}~, \\
I_2 & = \Delta_3 ~,\\
I_4 &=  \Delta_2\Delta_3  - \frac{c+6}{24}\Delta_3~,\\
I_5 &=\Delta _2^3+9 \Delta _3^2 - \left(\frac{c+8}{8}\right) \Delta _2^2 
+\frac{1}{192} (c+2) (c+15) \Delta _2-\frac{c (c+23) (7 c+30)}{96768} ~,\\
I_7 &=\Delta _2^4 +18 \Delta _3^2 \Delta_2 
-\frac{3}{4} (c+12) \Delta _3^2
\nonumber \\ 
&\hspace{2cm}-\frac{c+12}{6}\Delta _2^3 
+ \frac{1}{480} \left(5 c^2+127 c+594\right) \Delta _2^2  \nonumber\\ 
&\hspace{2cm} -\frac{(c+2) \left(10 c^2+387 c+3150\right)}{34560}\, \Delta _2
+\frac{c (5 c+186) \left(2 c^2+43 c+150\right)}{3317760} ~,\\
I_8&=\Delta _2^3\Delta _3 +3\Delta _3^3- \frac{(c+14)}{8}\Delta _2^2\Delta _3 +\frac{(c (5 c+148)+900)}{960}  \Delta _2\Delta _3
\nonumber\\
&\hspace{6cm}-\frac{(c+30) (c (35 c+604)+2940)}{483840}\Delta _3 ~,\\
I_{10}&= \Delta _2^4\Delta _3+ 6\Delta _2\Delta _3^3
-\frac{1}{6} (c+18) \Delta _2^3\Delta _3 
-\frac{1}{4} (c+18) \Delta _3^3\nonumber \\
&\hspace{2cm}+\frac{1}{96} (c+14) (c+24) \Delta _3 \Delta _2^2-\frac{(c+30) \left(7 c^2+215 c+1422\right)}{24192}\Delta _2 \Delta _3 \nonumber \\
& \hspace{6cm}+\frac{(c+30) \left(7 c^3+401 c^2+5844 c+26460\right) }{2322432}\Delta _3 ~,\\
I_{11}&= \Delta _2^6+\frac{135 }{2}\Delta _3^4+45 \Delta _3^2 \Delta _2^3
-\frac{1}{8} (c+20) \left(2 \Delta _2^5+45 \Delta _3^2\Delta_2^2\right)
\nonumber \\ 
&\hspace{1cm}
+\frac{\Delta _2}{192} \left(\Delta _2^3+9 \Delta _3^2\right) \left(5 c^2+211 c+2154\right) \nonumber \\
&\hspace{1cm}-\frac{35 c^3+2353 c^2+50514 c+383400}{10752} \Delta _3^2-\frac{\left(140 c^3+9445 c^2+202320 c+1233036\right) }{96768} \Delta _2^3\nonumber\\
&\hspace{1cm}+\frac{350 c^4+34061 c^3+1154072 c^2+14378340 c+50425200}{7741440} \Delta _2^2 \nonumber \\
&\hspace{1cm} -\frac{(c+2) \left(140 c^4+18577 c^3+865560 c^2+15423300 c+87318000\right) }{185794560}\Delta _2\nonumber \\
&\hspace{1cm}+\frac{c (7 c+470) \left(1820 c^4+216001 c^3+6997896 c^2+78367140 c+227026800\right)}{2434651914240}~.
 \end{align}
\end{subequations}
The constant prefactors are chosen in such a way that, in terms of the conformal field theory data, the eigenvalues of the even-dimension charges are normalized to be written as $\Delta_{2}^{n} + \ldots$, while the odd-dimension charges are normalized to be written as $\Delta_2^{n-1}\, \Delta_3 + \ldots$. This ensures that the quantum Boussinesq charges go over to the charges of the classical Boussinesq hierarchy in the large-$c$ limit. 

\bibliographystyle{JHEP}

\begin{thebibliography}{99}


\bibitem{Belavin:1984vu}
A.~A.~Belavin, A.~M.~Polyakov and A.~B.~Zamolodchikov,
``Infinite Conformal Symmetry in Two-Dimensional Quantum Field Theory,''
Nucl. Phys. B \textbf{241} (1984), 333-380
doi:10.1016/0550-3213(84)90052-X

\bibitem{Friedan:1980jm}
D.~H.~Friedan,
``Nonlinear Models in Two + Epsilon Dimensions,''
Annals Phys. \textbf{163} (1985), 318
doi:10.1016/0003-4916(85)90384-7

\bibitem{ZamolodchikovsBook}
A.~B.~Zamolodchikov, Al.~B.~Zamolodchikov, ``Conformal Field Theory and Critical Phenomena in Two Dimensional Systems," Physical Reviews, Edited by I.~M.~Khalatnikov, ISBN 10: 3718648636  ISBN 13: 9783718648634, Publisher: Routledge, 1989

\bibitem{DiFrancesco:1997nk}
P.~Di Francesco, P.~Mathieu and D.~Senechal,
``Conformal Field Theory,''
Springer-Verlag, 1997,
ISBN 978-0-387-94785-3, 978-1-4612-7475-9
doi:10.1007/978-1-4612-2256-9


\bibitem{Sasaki:1987mm}
R.~Sasaki and I.~Yamanaka,
``Virasoro Algebra, Vertex Operators, Quantum {Sine-Gordon} and Solvable Quantum Field Theories,''
Adv. Stud. Pure Math. \textbf{16} (1988), 271-296
RRK-87-3.

\bibitem{Eguchi:1989hs}
T.~Eguchi and S.~K.~Yang,
``Deformations of Conformal Field Theories and Soliton Equations,''
Phys. Lett. B \textbf{224} (1989), 373-378
doi:10.1016/0370-2693(89)91463-9

\bibitem{Kupershmidt:1989bf}
B.~A.~Kupershmidt and P.~Mathieu,
``Quantum Korteweg-de Vries Like Equations and Perturbed Conformal Field Theories,''
Phys. Lett. B \textbf{227} (1989), 245-250
doi:10.1016/S0370-2693(89)80030-9


\bibitem{Bazhanov:1994ft}
V.~V.~Bazhanov, S.~L.~Lukyanov and A.~B.~Zamolodchikov,
``Integrable structure of conformal field theory, quantum KdV theory and thermodynamic Bethe ansatz,''
Commun. Math. Phys. \textbf{177} (1996), 381-398
doi:10.1007/BF02101898
[arXiv:hep-th/9412229 [hep-th]].


\bibitem{Bazhanov:1996dr}
V.~V.~Bazhanov, S.~L.~Lukyanov and A.~B.~Zamolodchikov,
``Integrable structure of conformal field theory. 2. Q operator and DDV equation,''
Commun. Math. Phys. \textbf{190} (1997), 247-278
doi:10.1007/s002200050240
[arXiv:hep-th/9604044 [hep-th]].

\bibitem{Bazhanov:1996aq}
V.~V.~Bazhanov, S.~L.~Lukyanov and A.~B.~Zamolodchikov,
``Integrable quantum field theories in finite volume: Excited state energies,''
Nucl. Phys. B \textbf{489} (1997), 487-531
doi:10.1016/S0550-3213(97)00022-9
[arXiv:hep-th/9607099 [hep-th]].



\bibitem{Bazhanov:1998dq}
V.~V.~Bazhanov, S.~L.~Lukyanov and A.~B.~Zamolodchikov,
``Integrable structure of conformal field theory. 3. The Yang-Baxter relation,''
Commun. Math. Phys. \textbf{200} (1999), 297-324
doi:10.1007/s002200050531
[arXiv:hep-th/9805008 [hep-th]].

\bibitem{Zamolodchikov:1985wn}
A.~B.~Zamolodchikov,
``Infinite Additional Symmetries in Two-Dimensional Conformal Quantum Field Theory,''
Theor. Math. Phys. \textbf{65} (1985), 1205-1213
doi:10.1007/BF01036128

\bibitem{Lukyanov:1990tf}
S.~L.~Lukyanov and V.~A.~Fateev,
``Physics reviews: Additional symmetries and exactly soluble models in two-dimensional conformal field theory,''  Sov. Sci. Rev. A. Phys. 15 (1990) 1–117

\bibitem{Bazhanov:2001xm}
V.~V.~Bazhanov, A.~N.~Hibberd and S.~M.~Khoroshkin,
``Integrable structure of W(3) conformal field theory, quantum Boussinesq theory and boundary affine Toda theory,''
Nucl. Phys. B \textbf{622}, 475-547 (2002)
doi:10.1016/S0550-3213(01)00595-8
[arXiv:hep-th/0105177 [hep-th]].

\bibitem{Fioravanti:1995cq}
D.~Fioravanti, F.~Ravanini and M.~Stanishkov,
``Generalized KdV and quantum inverse scattering description of conformal minimal models,''
Phys. Lett. B \textbf{367} (1996), 113-120
doi:10.1016/0370-2693(95)01463-2
[arXiv:hep-th/9510047 [hep-th]].

\bibitem{Dorey:1998pt}
P.~Dorey and R.~Tateo,
``Anharmonic oscillators, the thermodynamic Bethe ansatz, and nonlinear integral equations,''
J. Phys. A \textbf{32} (1999), L419-L425
doi:10.1088/0305-4470/32/38/102
[arXiv:hep-th/9812211 [hep-th]].

\bibitem{Bazhanov:1998wj}
V.~V.~Bazhanov, S.~L.~Lukyanov and A.~B.~Zamolodchikov,
``Spectral determinants for Schrödinger equation and Q operators of conformal field theory,''
J. Statist. Phys. \textbf{102} (2001), 567-576
doi:10.1023/A:1004838616921
[arXiv:hep-th/9812247 [hep-th]].

\bibitem{Suzuki:1999rj}
J.~Suzuki,
``Anharmonic oscillators, spectral determinant and short exact sequence of U(q) (affine sl(2)),''
J. Phys. A \textbf{32} (1999), L183-L188
doi:10.1088/0305-4470/32/16/002
[arXiv:hep-th/9902053 [hep-th]].
 
 

\bibitem{Dorey:2000ma}
P.~Dorey, C.~Dunning and R.~Tateo,
``Differential equations for general SU(n) Bethe ansatz systems,''
J. Phys. A \textbf{33} (2000), 8427-8442
doi:10.1088/0305-4470/33/47/308
[arXiv:hep-th/0008039 [hep-th]].


\bibitem{Dorey:2006an}
P.~Dorey, C.~Dunning, D.~Masoero, J.~Suzuki and R.~Tateo,
``Pseudo-differential equations, and the Bethe ansatz for the classical Lie algebras,''
Nucl. Phys. B \textbf{772} (2007), 249-289
doi:10.1016/j.nuclphysb.2007.02.029
[arXiv:hep-th/0612298 [hep-th]].

\bibitem{Feigin:2007mr}
B.~Feigin and E.~Frenkel,
``Quantization of soliton systems and Langlands duality,''
in Exploring new structures and natural constructions in mathematical physics, volume 61 of Adv. Stud.
Pure Math., pages 185–274. Math. Soc. Japan, Tokyo, 2011.[arXiv:0705.2486 [math.QA]].

\bibitem{Bazhanov:2003ni}
V.~V.~Bazhanov, S.~L.~Lukyanov and A.~B.~Zamolodchikov,
``Higher level eigenvalues of Q operators and Schrödinger equation,''
Adv. Theor. Math. Phys. \textbf{7} (2003) no.4, 711-725
doi:10.4310/ATMP.2003.v7.n4.a4
[arXiv:hep-th/0307108 [hep-th]].

\bibitem{Masoero:2018rel}
D.~Masoero and A.~Raimondo,
``Opers for higher states of quantum KdV models,''
Commun. Math. Phys. \textbf{378} (2020) no.1, 1-74
doi:10.1007/s00220-020-03792-3
[arXiv:1812.00228 [math-ph]].

\bibitem{Masoero:2019wqf}
D.~Masoero and A.~Raimondo,
`Opers for higher states of the quantum Boussinesq model,''
doi:10.1007/978-3-030-57000-2\_5
[arXiv:1908.11559 [math-ph]].


\bibitem{Lukyanov:2010rn}
S.~L.~Lukyanov and A.~B.~Zamolodchikov,
``Quantum Sine(h)-Gordon Model and Classical Integrable Equations,''
JHEP \textbf{07} (2010), 008
doi:10.1007/JHEP07(2010)008
[arXiv:1003.5333 [math-ph]].


\bibitem{Lukyanov:2013wra}
S.~L.~Lukyanov,
``ODE/IM correspondence for the Fateev model,''
JHEP \textbf{12} (2013), 012
doi:10.1007/JHEP12(2013)012
[arXiv:1303.2566 [hep-th]].


\bibitem{Bazhanov:2013cua}
V.~V.~Bazhanov and S.~L.~Lukyanov,
``Integrable structure of Quantum Field Theory: Classical flat connections versus quantum stationary states,''
JHEP \textbf{09} (2014), 147
doi:10.1007/JHEP09(2014)147
[arXiv:1310.4390 [hep-th]].

\bibitem{Dorey:2007zx}
P.~Dorey, C.~Dunning and R.~Tateo,
``The ODE/IM Correspondence,''
J. Phys. A \textbf{40} (2007), R205
doi:10.1088/1751-8113/40/32/R01
[arXiv:hep-th/0703066 [hep-th]].

\bibitem{Negro:2017xwc}
S.~Negro,
`ODE/IM Correspondence in Toda Field Theories and Fermionic Basis in sin(h)-Gordon Model,''
[arXiv:1702.06657 [hep-th]].


\bibitem{Dorey:2019ngq}
P.~Dorey, C.~Dunning, S.~Negro and R.~Tateo,
`Geometric aspects of the ODE/IM correspondence,''
J. Phys. A \textbf{53} (2020) no.22, 223001
doi:10.1088/1751-8121/ab83c9
[arXiv:1911.13290 [hep-th]].

\bibitem{Sibuya}
Y.~Sibuya, ``Global Theory of a Second Order Linear Ordinary Differential Equation with a Polynomial Coefficient,'' 
North-Holland Publishing Company, 1975.

\bibitem{Voros1}
A.~Voros, ``Spectral Zeta Functions," Adv. Stud. Pure Math. {\bf 121} (1992), no. 3, 327–358.

\bibitem{Voros2}
A.~Voros, ``Exact quantization condition for anharmonic oscillators (in one dimension)," Journal of Physics A: Mathematical and General {\bf 27} (1994), no. 13, 4653.

\bibitem{Dorey:1999uk}
P.~Dorey and R.~Tateo,
``On the relation between Stokes multipliers and the T-Q systems of conformal field theory,''
Nucl. Phys. B \textbf{563} (1999), 573-602
[erratum: Nucl. Phys. B \textbf{603} (2001), 581-581]
doi:10.1016/S0550-3213(99)00609-4
[arXiv:hep-th/9906219 [hep-th]].



\bibitem{Dymarsky:2022dhi}
A.~Dymarsky, A.~Kakkar, K.~Pavlenko and S.~Sugishita,
``Spectrum of quantum KdV hierarchy in the semiclassical limit,''
JHEP \textbf{09} (2022), 169
doi:10.1007/JHEP09(2022)169
[arXiv:2208.01062 [hep-th]].

\bibitem{Maloney:2018hdg}
A.~Maloney, G.~S.~Ng, S.~F.~Ross and I.~Tsiares,
``Thermal Correlation Functions of KdV Charges in 2D CFT,''
JHEP \textbf{02} (2019), 044
doi:10.1007/JHEP02(2019)044
[arXiv:1810.11053 [hep-th]].


\bibitem{zhu:1990}
Y. Zhu,
  ``Vertex operator algebras, elliptic functions and modular forms," Ph.D. dissertation,
Yale Univ., 1990. https://api.semanticscholar.org/CorpusID:117186653

\bibitem{zhu:1996}
Y. Zhu, 
``Modular Invariance of Characters of Vertex Operator Algebras," Journal of the American Mathematical Society, Vol. 9, No. 1 (Jan., 1996), pp. 237-302


\bibitem{boussinesq1872theorie}
J.~Boussinesq,
``Th{\'e}orie des ondes et des remous qui se propagent le long d'un canal rectangulaire horizontal, en communiquant au liquide contenu dans ce canal des vitesses sensiblement pareilles de la surface au fond,''
Journal de math{\'e}matiques pures et appliqu{\'e}es (1872),
  volume 17 : 55-108.


\bibitem{Mckean:1978}
McKean, H. P. "Boussinesq’s equation as a Hamiltonian system." Adv. Math. Supp. Studies 3, no. 229 (1978): 217-226.

\bibitem{zakharov1973stochastization}
V.~E.~Zakharov,
``On stochastization of one-dimensional chains of nonlinear oscillators,''
Zh. Eksp. Teor. Fiz (1973),
Vol. 65

\bibitem{Ito:2023zdc}
K.~Ito and M.~Zhu,
``WKB analysis of the linear problem for modified affine Toda field equations,''
JHEP \textbf{08} (2023), 007
doi:10.1007/JHEP08(2023)007
[arXiv:2305.03283 [hep-th]].



\bibitem{Drinfeld:1984qv}
V.~G.~Drinfeld and V.~V.~Sokolov,
``Lie algebras and equations of Korteweg-de Vries type,''
J. Sov. Math. \textbf{30} (1984), 1975-2036
doi:10.1007/BF02105860




 
\bibitem{Ito:2013aea}
K.~Ito and C.~Locke,
``ODE/IM correspondence and modified affine Toda field equations,''
Nucl. Phys. B \textbf{885} (2014), 600-619
doi:10.1016/j.nuclphysb.2014.06.007
[arXiv:1312.6759 [hep-th]].

\bibitem{Adamopoulou:2014fca}
P.~Adamopoulou and C.~Dunning,
``Bethe Ansatz equations for the classical $A_n^{(1)}$ affine Toda field theories,''
J. Phys. A \textbf{47} (2014), 205205
doi:10.1088/1751-8113/47/20/205205
[arXiv:1401.1187 [math-ph]].

\bibitem{Dorey:1999pv}
P.~Dorey and R.~Tateo,
``Differential equations and integrable models: The SU(3) case,''
Nucl. Phys. B \textbf{571}, 583-606 (2000)
[erratum: Nucl. Phys. B \textbf{603}, 582-582 (2001)]
doi:10.1016/S0550-3213(99)00791-9
[arXiv:hep-th/9910102 [hep-th]].






\bibitem{Dorey:2004fk}
P.~Dorey, A.~Millican-Slater and R.~Tateo,
``Beyond the WKB approximation in PT-symmetric quantum mechanics,''
J. Phys. A \textbf{38}, 1305-1332 (2005)
doi:10.1088/0305-4470/38/6/010
[arXiv:hep-th/0410013 [hep-th]].


\bibitem{Langer:1937qr}
R.~E.~Langer,
``On the Connection Formulas and the Solutions of the Wave Equation,''
Phys. Rev. \textbf{51} (1937), 669-676
doi:10.1103/PhysRev.51.669



\bibitem{Ito:2021boh}
K.~Ito, T.~Kondo, K.~Kuroda and H.~Shu,
``WKB periods for higher order ODE and TBA equations,''
JHEP \textbf{10} (2021), 167
doi:10.1007/JHEP10(2021)167
[arXiv:2104.13680 [hep-th]].


\bibitem{Lacroix:2018fhf}
S.~Lacroix, B.~Vicedo and C.~Young,
``Affine Gaudin models and hypergeometric functions on affine opers,''
Adv. Math. \textbf{350} (2019), 486-546
doi:10.1016/j.aim.2019.04.032
[arXiv:1804.01480 [math.QA]].


\bibitem{Lacroix:2018itd}
S.~Lacroix, B.~Vicedo and C.~A.~S.~Young,
``Cubic hypergeometric integrals of motion in affine Gaudin models,''
Adv. Theor. Math. Phys. \textbf{24} (2020) no.1, 155-187
doi:10.4310/ATMP.2020.v24.n1.a5
[arXiv:1804.06751 [math.QA]].




\bibitem{KanekoZagier}
M.~Kaneko and D.~Zagier, ``A generalised Jacobi theta function and quasimodular
forms," Progress in Mathematics, Vol. {\bf 129}, 165 (Birkh¨auser, Boston, 1995).

\bibitem{Gaberdiel:2012yb}
M.~R.~Gaberdiel, T.~Hartman and K.~Jin,
``Higher Spin Black Holes from CFT,''
JHEP \textbf{04} (2012), 103
doi:10.1007/JHEP04(2012)103
[arXiv:1203.0015 [hep-th]].

\bibitem{Iles:2013jha}
N.~J.~Iles and G.~M.~T.~Watts,
``Characters of the $W_3$ algebra,''
JHEP \textbf{02} (2014), 009
doi:10.1007/JHEP02(2014)009
[arXiv:1307.3771 [hep-th]].

\bibitem{Iles:2014gra}
N.~J.~Iles and G.~M.~T.~Watts,
``Modular properties of characters of the W$_{3}$ algebra,''
JHEP \textbf{01} (2016), 089
doi:10.1007/JHEP01(2016)089
[arXiv:1411.4039 [hep-th]].


\bibitem{Maloney:2018yrz}
A.~Maloney, G.~S.~Ng, S.~F.~Ross and I.~Tsiares,
``Generalized Gibbs Ensemble and the Statistics of KdV Charges in 2D CFT,''
JHEP \textbf{03} (2019), 075
doi:10.1007/JHEP03(2019)075
[arXiv:1810.11054 [hep-th]].


\bibitem{Dymarsky:2018lhf}
A.~Dymarsky and K.~Pavlenko,
``Generalized Gibbs Ensemble of 2d CFTs at large central charge in the thermodynamic limit,''
JHEP \textbf{01} (2019), 098
doi:10.1007/JHEP01(2019)098
[arXiv:1810.11025 [hep-th]].


\bibitem{Dijkgraaf:1996iy}
R.~Dijkgraaf,
``Chiral deformations of conformal field theories,''
Nucl. Phys. B \textbf{493} (1997), 588-612
doi:10.1016/S0550-3213(97)00153-3
[arXiv:hep-th/9609022 [hep-th]].



\bibitem{Conti:2021xzr}
R.~Conti and D.~Masoero,
``On Solutions of the Bethe Ansatz for the Quantum KdV Model,''
Commun. Math. Phys. \textbf{402} (2023) no.1, 335-390
doi:10.1007/s00220-023-04728-3
[arXiv:2112.14625 [math-ph]].

\bibitem{Dymarsky:2019iny}
A.~Dymarsky, K.~Pavlenko and D.~Solovyev,
``Zero modes of local operators in 2d CFT on a cylinder,''
JHEP \textbf{07} (2020), 172
doi:10.1007/JHEP07(2020)172
[arXiv:1912.13444 [hep-th]].


\bibitem{Lashkari:2016vgj}
N.~Lashkari, A.~Dymarsky and H.~Liu,
``Eigenstate Thermalization Hypothesis in Conformal Field Theory,''
J. Stat. Mech. \textbf{1803} (2018) no.3, 033101
doi:10.1088/1742-5468/aab020
[arXiv:1610.00302 [hep-th]].


\bibitem{Dymarsky:2019etq}
A.~Dymarsky and K.~Pavlenko,
``Generalized Eigenstate Thermalization Hypothesis in 2D Conformal Field Theories,''
Phys. Rev. Lett. \textbf{123} (2019) no.11, 111602
doi:10.1103/PhysRevLett.123.111602
[arXiv:1903.03559 [hep-th]].

\bibitem{Dymarsky:2018iwx}
A.~Dymarsky and K.~Pavlenko,
``Exact generalized partition function of 2D CFTs at large central charge,''
JHEP \textbf{05} (2019), 077
doi:10.1007/JHEP05(2019)077
[arXiv:1812.05108 [hep-th]].



\bibitem{Handbooks}
  ``FunGrim: a symbolic library for special functions,'' arXiv:2003.06181 [cs.MS], \url{https://fungrim.org}.


\bibitem{Goddard:1989dp}
P.~Goddard, 
``Meromorphic Conformal Field Theory,'' in: Infinite Dimensional Lie Algebras and Lie Groups, edited by V.~Kac, page 556, World Scientific, Singapore, New Jersey, Hong Kong, DAMTP-89-01.

\bibitem{Gaberdiel:1999mc}
M.~R.~Gaberdiel,
``An Introduction to conformal field theory,''
Rept. Prog. Phys. \textbf{63} (2000), 607-667
doi:10.1088/0034-4885/63/4/203
[arXiv:hep-th/9910156 [hep-th]].
\bibitem{tuite}
M,~P,~Tuite,
  ``Modular Forms in Vertex Operator Algebras,'' \url{https://legacy.slmath.org/attachments/sgw/449/tuite.pdf}.

\bibitem{Mason:2009xt}
G.~Mason and M.~P.~Tuite,
``Vertex Operators and Modular Forms,''
[arXiv:0909.4460 [math.QA]].

\bibitem{Gaberdiel:1994fs}
M.~Gaberdiel,
``A General transformation formula for conformal fields,''
Phys. Lett. B \textbf{325} (1994), 366-370
doi:10.1016/0370-2693(94)90026-4
[arXiv:hep-th/9401166 [hep-th]].


\end{thebibliography}

\end{document}